# An X-ray detection of star formation in a highly magnified giant arc

M. B. Bayliss[1,2]*, M. McDonald[1], K. Sharon[3], M. D. Gladders[4,5], M. Florian[6], J. Chisholm[7], H. Dahle[8], G. Mahler[3], R. Paterno-Mahler[9], J. R. Rigby[6], E. Rivera-Thorsen[8], K. E. Whitaker[10,11,12], S. Allen[13,14,15], B. A. Benson[4,5,16], L. E. Bleem[17], M. Brodwin[18], R. E. A. Canning[13,14], I. Chiu[19], J. Hlavacek-Larrondo[20], G. Khullar[4,5], C. Reichardt[21] and J. D. Vieira[22]

In the past decade, our understanding of how stars and galaxies formed during the first 5 billion years after the Big Bang has been revolutionized by observations that leverage gravitational lensing by intervening masses, which act as natural cosmic telescopes to magnify background sources. Previous studies have harnessed this effect to probe the distant Universe at ultraviolet, optical, infrared and millimetre wavelengths[1–6]. However, strong-lensing studies of young, star-forming galaxies have never extended into X-ray wavelengths, which uniquely trace high-energy phenomena. Here, we report an X-ray detection of star formation in a highly magnified, strongly lensed galaxy. This lensed galaxy, seen during the first third of the history of the Universe, is a low-mass, low-metallicity starburst with elevated X-ray emission, and is a likely analogue to the first generation of galaxies. Our measurements yield insight into the role that X-ray emission from stellar populations in the first generation of galaxies may play in reionizing the Universe. This observation paves the way for future strong-lensing-assisted X-ray studies of distant galaxies reaching orders of magnitude below the detection limits of current deep fields, and previews the depths that will be attainable with future X-ray observatories.

The massive galaxy cluster, SPT-CLJ2344-4243 (the Phoenix cluster), acts as a gravitational lens, magnifying our view of a background star-forming galaxy. The background galaxy is at a redshift $z = 1.5244$, such that we are observing it at a cosmic age of 4.2 billion years after the Big Bang (using the current Planck cosmological parameter values[7] for the Hubble constant, $H_0 = 67.4$ km s$^{-1}$ Mpc$^{-1}$, the matter density, $\Omega_{\rm m} = 0.315$, and density of vacuum energy, $\Omega_\Lambda = 0.685$). The foreground lens, SPT-CLJ2344-4243, was discovered in a millimetre-wave survey of 2,500 deg$^2$ of the southern sky by the South Pole Telescope (SPT)[8], has a measured redshift of $z = 0.596$, and has very dense core[9] that likely contributes to its efficacy as a natural gravitational telescope. The highly magnified, lensed galaxy was discovered serendipitously in optical imaging data, where it appears as a thin giant arc extending approximately 12 arcsec (″) across the sky. Using follow-up optical imaging from the Hubble Space Telescope, we confirmed that the giant arc is formed by gravitational lensing of a faint star-forming background galaxy. Remarkably, a deep (~600 kilosecond (ks)) observation taken with the Chandra X-ray Observatory for the purpose of measuring the X-ray emission from SPT-CLJ2344-4243 also revealed the presence of X-ray emission from each pair of merging images that make up the giant arc (Fig. 1). We model and subtract the spatially extended foreground X-ray emission from the cluster, resulting in 30.6 ± 6.3 net, background-subtracted X-ray counts from the giant arc in the 0.5–7 keV band, and a 5.3σ detection significance.

We obtained near-infrared (NIR) spectra at three different positions along the X-ray-emitting arc with the Folded-port InfraRed Echellette (FIRE) instrument on the Magellan-I telescope; these spectra indicate the presence of multiple rest-frame optical nebular emission lines at a redshift of $z = 1.5244$. The similarity of the spectra taken at different positions along the arc confirm that the arc consists of two merging images with mirror symmetry. The lensed galaxy spectrum contains strong optical emission lines from a variety of different elements and ions (Hα, Hβ, [O II], [O III] and [N II]). The relative strengths of these lines reveal the physical properties of the ionized nebular gas in the lensed galaxy[10,11]. In particular, the [N II]/Hα and [O III]/Hβ ratios are typical of those observed in star-forming galaxies and appear inconsistent with the expectations for an active galactic nucleus (AGN), and hence demonstrate that the observed X-ray emission is from ongoing star formation (see Methods for more details).

Having confirmed that the giant arc is a single strongly lensed star-forming galaxy, we created a model reconstruction of the gravitational lensing due to the massive galaxy cluster. FIRE spectra of seven multiply-imaged, lensed background galaxies with images extending from the cluster core to beyond the giant arc, along with

[1]Kavli Institute for Astrophysics and Space Research, Massachusetts Institute of Technology, Cambridge, MA, USA. [2]Department of Physics, University of Cincinnati, Cincinnati, OH, USA. [3]Department of Astronomy, University of Michigan, Ann Arbor, MI, USA. [4]Department of Astronomy and Astrophysics, University of Chicago, Chicago, IL, USA. [5]Kavli Institute for Cosmological Physics, University of Chicago, Chicago, IL, USA. [6]Observational Cosmology Lab, Goddard Space Flight Center, Greenbelt, MD, USA. [7]Department of Astronomy and Astrophysics, University of California, Santa Cruz, Santa Cruz, CA, USA. [8]Institute of Theoretical Astrophysics, University of Oslo, Oslo, Oslo, Norway. [9]Department of Physics and Astronomy, University of California, Irvine, Irvine, CA, USA. [10]Department of Astronomy, University of Massachusetts, Amherst, MA, USA. [11]Department of Physics, University of Connecticut, Storrs, CT, USA. [12]Cosmic Dawn Center (DAWN), Copenhagen, Denmark. [13]Kavli Institute for Particle Astrophysics and Cosmology, Stanford University, Stanford, CA, USA. [14]Department of Physics, Stanford University, Stanford, CA, USA. [15]SLAC National Accelerator Laboratory, Menlo Park, CA, USA. [16]Fermi National Accelerator Laboratory, Batavia, IL, USA. [17]Argonne National Laboratory, High-Energy Physics Division, Argonne, IL, USA. [18]Department of Physics and Astronomy, University of Missouri, Kansas City, MO, USA. [19]Academia Sinica Institute of Astronomy and Astrophysics, Taipei, Taiwan. [20]Department of Physics, University of Montreal, Montreal, Quebec, Canada. [21]School of Physics, University of Melbourne, Parkville, Victoria, Australia. [22]Department of Astronomy and Department of Physics, University of Illinois, Urbana, IL, USA. *e-mail: matthew.bayliss@uc.edu

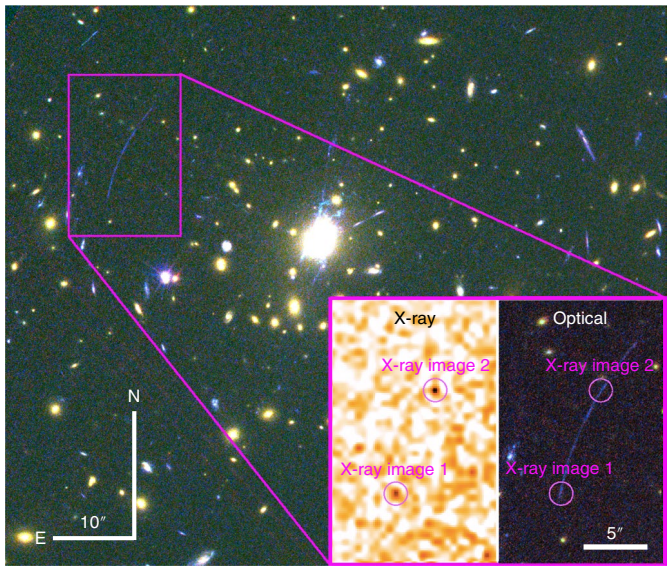

**Fig. 1 | False-colour and X-ray images of the giant arc in SPT-CLJ2344-4243.** The X-ray-emitting giant arc is shown relative to the centre of the foreground lensing galaxy cluster in a false-colour image at optical wavelengths. The inset shows Chandra X-ray 0.5–7 keV (left) and Hubble optical (right) images of the giant arc at a scale 1.5 times larger. The optical colours here are given by Hubble imaging data in the F850LP (red), F775W (green) and F475W (blue) filters. Two magenta circles indicate the locations of the X-ray emission from the giant arc in both inset panels. The lensing geometry of the giant arc is a pair of merging images, where the lower and upper halves of the arc are each a single image with mirror symmetry.

**Table 1 | Properties of the X-ray arc**

| Observed X-ray luminosity | (erg s$^{-1}$) |
|---|---|
| $L_{X,0.5-8}$ | $8.3 \pm 1.7 \times 10^{42}$ |
| $L_{X,2-10}$ | $6.2 \pm 1.3 \times 10^{42}$ |
| **Observed Hubble photometry** | **(AB magnitudes)** |
| $m_{F475W}$ | $23.66 \pm 0.04$ |
| $m_{F775W}$ | $23.41 \pm 0.09$ |
| $m_{F850LP}$ | $23.57 \pm 0.36$ |
| **Observed Spitzer photometry** | **(μJy)** |
| 3.6 μm | <13 |
| 4.5 μm | <23 |
| **Lensed galaxy intrinsic properties** | |
| $M_{UV,AB}$ | $-16.3 \pm 0.4$ |
| $L_{X,0.5-8}$ (erg s$^{-1}$) | $6.5 \pm 2.3 \times 10^{40}$ |
| $L_{X,2-10}$ (erg s$^{-1}$) | $4.7 \pm 1.7 \times 10^{40}$ |
| $\log(M_\star/M_\odot)$ | <8.0 |
| $E(B-V)_{gas}$ | <0.15 |
| $Z/Z_\odot$ | $0.25^{+0.15}_{-0.1}$ |
| SFR$_{H\alpha}$ ($M_\odot$ yr$^{-1}$)[a] | $0.8^{+0.4}_{-0.3}$ |
| SFR$_{UV}$ ($M_\odot$ yr$^{-1}$)[a] | $0.5^{+0.5}_{-0.3}$ |
| Electron density, $n_e$ (cm$^{-3}$) | $1,000 \pm 200$ |

Uncertainties reported are 1$\sigma$ (68% confidence interval). [a]Reported SFRs include measurement and extinction uncertainties.

the giant arc itself, robustly constrain the model (see Methods). From the lens model, the best-fit magnification of the giant arc is $65 \pm 20$. We also use the lens model to reconstruct de-lensed images of the giant arc in the source plane, and find that the source is an irregular blue galaxy composed of two star-forming clumps, each less than a kiloparsec in diameter, of similar brightness at ~1,900 Å in the rest frame and separated by ~500 pc in projection. The X-ray emission from the giant arc is associated with one of the two star-forming clumps and has an intrinsic luminosity in the rest-frame 2–10 keV band of $L_{X,2-10} = 4.7 \pm 1.7 \times 10^{40}$ erg s$^{-1}$ ($L_{X,0.5-8} = 6.5 \pm 2.3 \times 10^{40}$ erg s$^{-1}$ in the 0.5–8 keV band), where the uncertainties reported here include both measurement and lens model (magnification) uncertainties as described in the Methods. The X-ray emission per unit stellar mass from star formation peaks in ≲30-Myr-old stellar populations[12], suggesting that the X-ray-emitting UV-bright clump in this lensed galaxy is most likely an extremely young star-forming region.

We measured the lensed galaxy to have a star-formation rate (SFR) between SFR = $0.8^{+0.4}_{-0.3}$ $M_\odot$ yr$^{-1}$ using H$\alpha$ emission and accounting for potential extinction due to intervening dust. We also placed an upper limit on the stellar mass of the galaxy of $M_\star < 1.0 \times 10^8 M_\odot$ using rest-frame NIR photometry. The SFR and mass constraints imply a specific star-formation rate (sSFR) >8×10$^{-9}$ yr$^{-1}$, confirming the lensed galaxy to be a typical low-mass (dwarf) star-forming galaxy, with its luminosity likely dominated by young stars. All measured properties of the giant arc are given in Table 1. The only X-ray detections to date of star formation in individual galaxies are either in the local Universe[13–17] or in the deepest X-ray deep fields[18–20]. Blind stacking analyses of large samples of galaxies[21,22] have also yielded relatively low signal-to-noise measurements of the average X-ray emission from galaxies in broad redshift bins out to $z \sim 5$. The galaxy that is lensed into the X-ray arc in

SPT-CLJ2344-4243 is among the most distant individual galaxies in which ongoing star formation has been detected in X-rays. This X-ray-detected giant arc is much fainter than the few detections at comparable redshifts, with an intrinsic X-ray luminosity that is more than an order of magnitude below the typical $z > 1.5$ galaxies with X-rays detected in deep fields (Fig. 2).

High-sSFR galaxies contain young stellar populations, with emission dominated by hot, massive stars. X-ray observations can directly detect the subset of these massive stars that are formed in gravitationally bound binaries[23]; when one star in the binary pair collapses into a black hole or neutron star, it can accrete material from the companion massive star in what is called a high-mass X-ray binary (HMXB). Local studies point to a correlation between $L_X$ and SFR that reflects the direct physical relationship between the rate at which a galaxy is forming stars and the resulting population of HMXBs[18,23,24]. Star-forming galaxies dominated by young (age ≲30 Myr), low-metallicity stellar populations (that is, analogues of Lyman break galaxies and 'Green Pea' galaxies) follow a different scaling relation, and have larger $L_X$ at a given SFR[14,15,17,21,25]. The X-ray-emitting lensed galaxy in SPT-CLJ2344-4243 has observed rest-frame $L_X$-to-SFR ratio of $\log_{10}(L_{X,0.5-8}/SFR) = 40.91 \pm 0.25$. We note that the ratio of $L_X$ to SFR is, to first order, insensitive to uncertainties in the magnification. This observed ratio is 20 times higher than expected from the constant scaling relation of $\log_{10}(L_{X,0.5-8}/SFR) = 39.6$ from fitting to local and deep-field galaxy samples[26] (Fig. 3). We can also compare this source to empirical scaling relations and models that account for evolution in $L_X$/SFR as a function of metallicity[12,15,27]. For the metallicity of our source the empirical scaling relations predict $\log_{10}(L_{X,0.5-8}/SFR) = 39.75 \pm 0.34$, while the models predict $\log_{10}(L_{X,0.5-8}/SFR) = 40.1$; both of these predictions are ~6–14 times below the measured ratio in our strongly lensed dwarf galaxy, and the observed discrepancy is moderately (2.1$\sigma$) significant. Our measurement directly supports the idea that starbursting galaxies can generate substantially more X-ray emission from their stellar populations than more typical 'main sequence' star-forming galaxies, and that factors beyond SFR and metallicity—such as stellar population age[12] and sSFR[14]—are important for explaining the X-ray emission associated with star-forming galaxies.

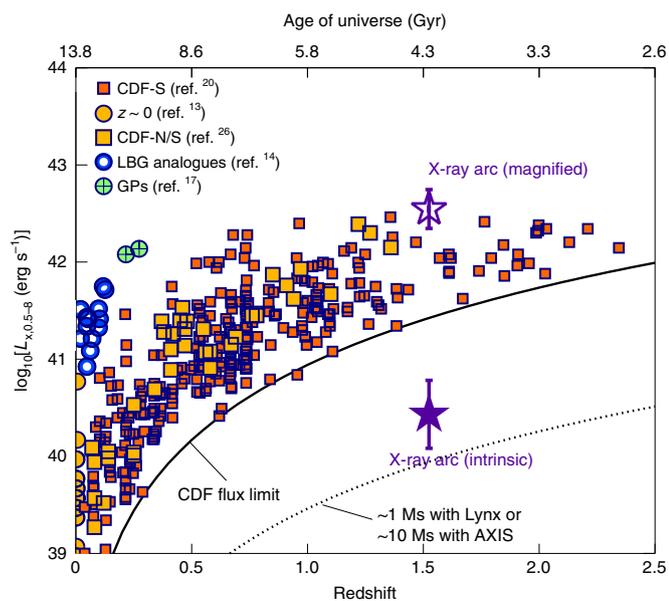

**Fig. 2 | Samples of star-forming galaxies with measured X-ray luminosities.** Here we show the observed X-ray luminosity versus redshift for X-ray-detected non-AGN galaxies, including local Lyman break galaxy analogues[14–16] (LBG analogs; blue open circles), 'Green Pea' extreme emission line galaxies[17] (GPs; green filled circles with crosshairs), and samples of 'normal' star-forming galaxies in the local Universe and in the northern and southern Chandra deep fields[13,18,20] (CDF-N/CDF-S; orange filled squares and red filled squares). The orange filled circles and squares identify galaxies used to fit a universal SFR to $L_X$ scaling relation[26]. Sources from the literature are plotted without error bars to make the plot easier to read; the literature uncertainties in $L_X$ are all in the range of ~5–20%. The X-ray arc is plotted as an open star for the observed (apparent) luminosity with the $1\sigma$ measurement uncertainty, and as a filled star for the true (intrinsic), lensing-corrected luminosity with the $1\sigma$ measurement plus magnification uncertainty. The lensing amplification has allowed us to detect X-ray emission from a dwarf star-forming galaxy ($z = 1.524$) with an intrinsic, lensing-corrected luminosity that existing Chandra deep fields (CDFs; solid black line) can only probe out to $z \sim 0.5$. Approximate depths for hypothetical deep fields from two proposed future X-ray observatories, AXIS and Lynx, are also plotted as the dotted black line.

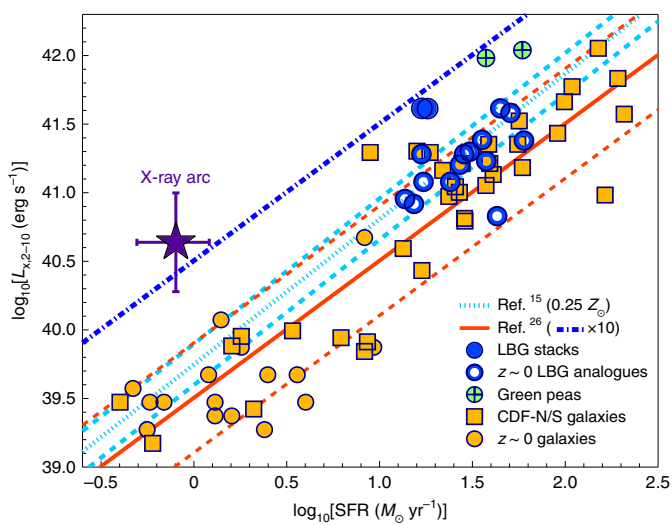

**Fig. 3 | Relationship between the X-ray luminosity and the SFR.** This figure shows the scaling between the X-ray luminosity and the SFR for galaxies not hosting an AGN. The samples of galaxies plotted are the same as in Fig. 2. For generic galaxy samples, the scaling between X-ray luminosity and the SFR follows a broad but consistent relationship spanning three decades in $L_X$ and the SFR[26] (solid red line with dispersion given by dashed red lines). We also show the best-fit relation at the measured metallicity, $Z = 0.25\,Z_\odot$, of our X-ray-emitting arc (cyan dotted line with dispersion given by dashed cyan lines) for published $L_X$-to-SFR scaling relations that allow for metallicity dependence[12,15]. For comparison we also plot several samples of galaxies from the literature with elevated $L_X$ at fixed SFR; these include Lyman break galaxy analogues[14–16] (LBG analogues; open blue circles); stacked high-redshift Lyman break galaxies (LBGs) in two bins centred on $z \sim 1.5$ and $z \sim 3$ from deep fields[21] (blue filled circles); and individually detected $z \sim 0.2$–0.3 extreme emission line galaxies[17] (so-called Green Peas; green filled circles with crosshairs). Sources from the literature are plotted without error bars to make the plot easier to read; the literature uncertainties in $L_X$ are all in the range of ~5–20%, and ~10–30% in the SFR. The X-ray-detected arc is plotted as a purple filled star with $1\sigma$ error bars that reflect both measurement and magnification uncertainties. All SFR values here are computed (or corrected) for a Salpeter IMF.

This elevated X-ray luminosity reflects a phase in the life cycle of star-forming galaxies during which HMXBs are present in large numbers. High-mass stellar binaries are thought to be important, if short-lived, contributors to high-energy emission in all galaxies that are dominated by young stellar populations, a stage through which all galaxies pass at some point in their evolutionary history. The X-ray-emitting giant arc in SPT-CLJ2344-4243 is a potential analogue of the first generation of galaxies that contributed to reionizing the Universe. Specifically, X-ray emission from young galaxies is likely an important contributor to the ionizing radiation budget, driven by emission from HMXB systems in nascent star-forming galaxies[28–30]. Young stellar populations (age $\lesssim 30$ Myr) may also play a key role in clearing out the interstellar medium, allowing ionizing radiation to escape galaxies, by generating powerful winds from HMXBs that can inject substantial mechanical power into their local environment[25]. Understanding the relationship between low-mass young star-forming galaxies and their HMXB populations is, therefore, crucial for understanding the physics of star formation across cosmic time and the reionization of the Universe by the first generation of stars and galaxies. In this lensed galaxy we are observing a typical low-mass, star-forming galaxy during the first third of the lifetime of the Universe, demonstrating how X-ray observations assisted by magnification from gravitational lensing enables studies that address the physical relationship between star formation and HMXB populations.

Our detection of a strongly lensed giant arc in the X-ray is an important first step that opens a new observational window into the formation of massive stars and HMXBs. This work demonstrates that X-ray facilities can be used in concert with strong lensing to push the limits of current X-ray telescopes and significantly improve our understanding of high-energy astrophysical phenomena. The combination of a deep Chandra exposure and high amplification by the foreground lensing potential produces an X-ray image of this distant galaxy at a depth equivalent to a ~1.3 year (40 megasecond (Ms)) Chandra exposure. It is also important to note that this detection was discovered serendipitously, and that while the X-ray arc in SPT-CLJ2344-4243 is highly magnified, it is also intrinsically very faint. Targeted observations of the brightest, most highly magnified giant arcs would enable much higher signal-to-noise X-ray detections with Chandra, with the potential to construct samples of X-ray-detected star-forming galaxies at high redshift in the near term. Furthermore, the next generation of X-ray observatories currently in development will be orders of magnitude more sensitive than Chandra. This discovery previews the kind of measurement that future missions will be capable of making en masse. An unlensed analogue of this lensed galaxy would require a ~2 month (4 Ms) integration—equivalent to a deep field—with the NASA Probe-class mission concept design for the Advanced X-ray Imaging Satellite (AXIS), and a ~5 day (~0.4 Ms) exposure with the proposed Chandra successor mission concept, Lynx. The combination of strong lensing with the sensitivity of proposed future X-ray missions would enable detailed studies of the brightest, most highly magnified star-forming galaxies, as well as ultra-deep searches for X-ray emission from galaxies out to $z \sim 10$. The former will spatially resolve X-ray emission from individual, distinct star-forming regions—and thereby link HMXB populations with the fundamental physical scales (that is, sub-galactic) on which stars formed in the distant Universe—while the latter will provide a powerful tool for studying the reionization history of the Universe.

## Methods

**Chandra X-ray Observatory data.** X-ray data for SPT-CLJ2344-4243 was obtained with Chandra ACIS-I over a series of programs in Cycle 12 (PI: Garmire, OBSID: 13401), Cycle 15 (PI: McDonald, OBSID: 16135, 16545), and Cycle 18 (PI: McDonald, OBSID: 19581, 19582, 19583, 20630, 20631, 20634, 20635, 20636, 20797). In total, this galaxy cluster was observed for a total of 551 ks, yielding roughly 300,000 counts in the 0.5–7.0 keV band. All Chandra data were first reprocessed using CIAO v4.10 and CALDB v4.8.0. Flares were identified following the procedure outlined by the calibration team and described online at http://cxc.harvard.edu/ciao/threads/flare/, using the 2.3–7.3 keV bandpass, time steps of 519.6 s and 259.8 s, a threshold of 2.5$\sigma$ and a minimum length of 3 time bins. A merged, exposure-corrected image was generated using the CIAO 'merge_obs' script, covering the broad energy range 0.5–7.0 keV. The bandpass was chosen to span the full sensitivity range of ACIS-I, while avoiding the high particle backgrounds at $E > 7.0$ keV. Point sources were identified on separate merged images in the 0.7–2.0 and 2.0–7.0 keV bands, using the WVDECOMP tool in the ZHTOOLS package[31]. The resulting list of point sources was used to generate a mask, which was applied to the broadband image.

The foreground emission from the galaxy cluster has a surface brightness profile that is well characterized out to large radius, with more than 200,000 total X-ray counts in the 0.5–7 keV channel (Supplementary Fig. 1). At the location of the giant arc, the cluster emission produces an average of $2.6 \pm 0.2$ counts per pixel in this energy band. This average level is very precisely measured, but on a per pixel basis the cluster emission is dominated by Poisson statistics. We fit the surface brightness of the X-ray-emitting intracluster medium (ICM) in SHERPA, using two two-dimensional (2D) beta functions, following the same methodology that we have previously applied to SPT-CLJ2344-4243[32]. The emission in the inner 3″ is dominated by a central AGN that resides at the core of the cluster, and was masked out before fitting to allow accurate modelling of the intracluster emission. The centres, ellipticities and position angles of the two models were tied, with the remaining parameters being allowed to float. The resulting best-fit model was subtracted from the broadband image, yielding an image free of X-ray emission from the ICM of the foreground galaxy cluster ('ICM-free'; Supplementary Fig. 2). The Chandra aim point was the core of the foreground lensing cluster, and the X-ray emission from the giant arc is at a cluster-centric radius of ~30″, well within the region where the Chandra point spread function (PSF) is at/near its minimum value.

The X-ray emission from the giant arc is visible as two point source excesses on top of the cluster background level (Supplementary Fig. 1, inset), as well as in the ICM-free subtracted image (Supplementary Fig. 2). We note that these two point sources coincide with the ends of the giant arc to the precision of the astrometry (0.3″). We describe the precision of the relative astrometric alignment of the Chandra and Hubble Space Telescope data below in the section 'Hubble Space Telescope imaging'. The two X-ray sources along the arc also have the expected mirror image symmetry appropriate to merging image pairs produced by strong gravitational lensing. The point source nature of the X-ray emission implies that it is localized to a physical region with a diameter no larger than ~400 pc in size. This size constraint is computed from the Chandra PSF and the strong-lensing magnification (strong-lens modelling is described in detail in the section 'Gravitational lens modelling' below), and is consistent with typical star-forming regions, such that the X-ray emission could easily result from a population of HMXB sources in a large star-forming complex in the lensed galaxy and still appear unresolved. It is, therefore, unclear how many distinct X-ray-emitting binary systems might be contributing to the observed X-ray flux. In the section 'The origin of the observed X-ray emission' below, we also explore the possibility that the two X-ray sources along the arc are the result of chance projection with background X-ray AGN, and find that the probability of such an occurrence is exceedingly small (<1 in 10$^7$).

Within an aperture of radius 0.5″ centred on each of the two point sources we measured a total of 19.2 and 19.1 counts from the north and south X-ray images, respectively. Within these apertures the modelled foreground cluster emission is 8.2 counts, so that the two X-ray images are 11.0 and 10.9 counts (0.5–7 keV band) above the background level. Given the background counts within the aperture, the background noise level is $\sigma_{bkgd} = \sqrt{8.2} = 2.86$, and the two X-ray sources along the arc are each independently detected at 3.8$\sigma$ significance above the background. The Chandra/ACIS pixel scale (0.49″) is approximately equal to the full-width at half-maximum of the PSF, so that the majority of the counts from each unresolved X-ray image of the lensed galaxy falls into a single pixel. We can, therefore, also examine the individual pixel-count statistics to quantify the significance of the X-ray detection of the giant arc. An annulus within which the foreground cluster emission is at the same level as it is at the location of the giant arc contains ~1,400 Chandra pixels, and the two highest pixel values in that annulus are the two brightest pixels that are coincident with the giant arc (that is, the central pixels of the two X-ray images). In the cluster-subtracted image, these two brightest pixels coincident with the giant arc are ~4$\sigma$ and ~6$\sigma$ significant, while the range of expected noise fluctuations in the annulus would be expected to include ~10 3$\sigma$ events, and only ~0.1 4$\sigma$ events.

We measured the precise flux from the giant arc using the ICM-free beta-model subtracted image. The flux from each image is strongly peaked in a central pixel, consistent with the Chandra PSF, and so we treated the two X-ray images as unresolved (point) sources. We computed the counts from each X-ray image using circular apertures and an algorithm based on the open-source IRAF/PyRAF aperture photometry APPHOT package. We measured total X-ray counts from the giant arc using two apertures with radii of 0.5″ and 1.0″, representing the 77% and 92% enclosed energy radii, respectively. In the 0.5″ and 1.0″ apertures we measured statistically identical fluxes, and so we used the higher signal-to-noise measurements ($r = 0.5''$). The final enclosed-energy corrected net count levels are $14.9 \pm 4.4$ and $15.7 \pm 4.5$ from the north and south X-ray images, respectively. The total count yield from the two mirrored images that form the giant arc is therefore $30.6 \pm 6.3$ in the 0.5–7 keV band, which we converted into apparent luminosities in the rest-frame 0.5–8 and 2–10 keV bands, assuming a typical HMXB spectral index of 1.7. The resulting total apparent luminosities are $L_{X,0.5-8} = 8.3 \pm 1.7 \times 10^{42}$ erg s$^{-1}$ and $L_{X,2-10} = 6.1 \pm 1.3 \times 10^{42}$ erg s$^{-1}$. Using our strong-lensing model (described in the section 'Gravitational lens modelling' below) for the foreground lens we estimated the magnification acting on both images of the X-ray emission from the giant arc to be $65 \pm 20$. We used the combined (averaged) emission from the two X-ray images of the galaxy to optimize the signal-to-noise ratio of the X-ray luminosity measurement. This means converting from the total apparent X-ray luminosity by dividing by the magnification factor acting on each arc, and then dividing by a factor of two to account for the fact that we are measuring the total signal from two images of the same intrinsic source. The resulting intrinsic rest-frame X-ray luminosities are $L_{X,0.5-8} = 6.5 \pm 2.3 \times 10^{40}$ erg s$^{-1}$ and $L_{X,2-10} = 4.7 \pm 1.7 \times 10^{40}$ erg s$^{-1}$ for the lensed galaxy in the 0.5–8 and 2–10 keV bands, respectively. The low number of counts precludes a detailed analysis of the X-ray spectrum of the lensed source, but we did measure its X-ray hardness ratio, defined as (hard − soft)/(hard + soft), using a 'soft' band of 0.7–2.0 keV and a 'hard' band of 2–7 keV in the observed frame. The hardness of the lensed source is $0.2 \pm 0.3$, which is consistent with measured hardness ratios of the $z \gtrsim 1$ star-forming galaxies detected in X-ray deep fields[20,33]. The X-ray luminosity of the giant arc is listed in Table 1, along with its other observable properties.

We also examined the locations of candidate counter-images of the X-ray arc that were predicted by the strong-lens model (described in detail in the section 'Gravitational lens modelling' below). There are two candidate counter-images predicted, and we find no evidence for excess emission above that of the foreground galaxy cluster. This result is expected due to the counter-images being much less magnified than the giant arc. At the location of the two predicted counter-images the magnifications are 6 and 4.6, which means that the X-ray count rate of these counter-images should be approximately 10–13 times lower than the giant arc, or on the order of 1 total count expected per counter-image (that is, undetectable).

**Magellan/FIRE infrared spectroscopy.** We observed the giant arc, as well as several other lensed, multiply-imaged background sources, with the Folded-port InfraRed Echellette (FIRE) spectrograph at the Magellan-I Baade telescope on 27–28 August and 20–21 September 2018. FIRE delivers spectra with a resolution of $R = 4,000$ and wavelength range of 0.82–2.5 µm in a single-object cross-dispersed set-up with the 0.75″ wide slit[34]. Observations of the giant arc were taken with three different slit positions and multiple rotation angles to guard against artefacts masquerading as emission lines. On 27 August we used a 0.6″ wide slit, resulting in higher resolution but lower throughput. All other observations were performed with the 0.75″ slit, which delivers spectra broadened by an instrumental velocity width of 63 km s$^{-1}$. The 0.6″ slit spectra have an instrumental line width of 50 km s$^{-1}$. Slits were placed on three different positions along the giant arc, as shown in Supplementary Fig. 3. We also placed the FIRE slit on 11 additional sources within the field of SPT-CLJ2344-4243 in an effort to identify additional multiply-imaged background sources to constrain a strong-lensing model of the foreground galaxy cluster.

In the raw 2D science frames, we identified emission lines for a total of 11 sources, including the giant arc. None of the science targets exhibited smooth continuum emission. For emission line sources, the FIRE reduction pipeline (FIREHOSE) takes user-supplied positions of visually identified emission lines to fit a model for the source trace. Given emission line positions and a trace model, FIREHOSE extracts object spectra by jointly fitting the source trace along with the 2D sky spectrum using the regions along the slit that are not illuminated by the source. The extracted source spectra were flux calibrated using spectra of A0V stars taken at similar airmass, and coincident to within 1 h, of each science frame. Telluric absorption was corrected[35] using the xtellcor procedure as a part of the spextool pipeline[36], which is called as a part of the FIREHOSE reduction process.

We measured cosmological redshifts for each source with an extracted FIRE spectrum by identifying families of nebular emission lines—generally, some combination of Hα, Hβ, [O III] at $\lambda\lambda 4960, 5008$ Å, and [O II] at 3727,3729 ÅÅ— and fitting a Gaussian profile to each emission line. The mean redshift for each spectrum was estimated as the average of the individual line redshifts, and the uncertainty was computed by adding in quadrature the uncertainties in the individual line centroids from each Gaussian profile, the uncertainty in the wavelength solution (never the dominant contributor), and the scatter in the measured redshifts of the individual emission lines.

The FIRE spectra resulting from the three different slit positions along the arc all exhibited strong rest-frame optical emission lines from hydrogen (Hα at 6564 Å and Hβ at 4862 Å) and oxygen ([O III] at 5008 Å and 4960 Å; and [O II] at 3727 Å and 3729 Å) at a redshift of $z = 1.5244$, confirming that the giant arc is a multiply-imaged background galaxy. We combined the FIRE spectra from the three different slit positions into a single extracted spectrum to achieve the best possible signal-to-noise ratio for the purpose of measuring the physical properties of the galaxy. The combined spectra include data taken with the 0.6″ and 0.75″ slits, and have a velocity resolution of 63 km s$^{-1}$, set by the 0.75″ slit data. In the combined spectrum we also weakly detected emission from nitrogen ([N II] at 6585 Å). Supplementary Fig. 3 shows the detected emission lines from the final combined FIRE spectrum of the giant arc.

We measured the total fluxes in each emission line in the giant arc spectrum by summing the flux of each emission line over a velocity interval of 120 km s$^{-1}$ centred on the Gaussian best-fit centroid of each emission line. The 120 km s$^{-1}$ interval was selected to encompass all of the appreciable signal from each line, which have measured velocity full-widths at half-maximums of 60–80 km s$^{-1}$ (consistent with the instrumental velocity spread) but are also narrow enough to avoid contaminating flux from residuals from nearby sky lines. The observed rest-frame emission line fluxes, uncorrected for extinction, are reported in Supplementary Table 1. The line flux measurement errors result from a combination of statistical uncertainties in the data due to background noise and sky subtraction, as well as additional systematic uncertainty in the time-variable telluric absorption features that overlap with the emission lines. This systematic uncertainty is at the ~10% level for all lines, as estimated from observed variations in the strength of these atmospheric absorption lines even in sequential exposures of the same standard star.

**Hubble Space Telescope imaging.** We used Hubble imaging in three optical bands, F475W, F775W and F850LP, that were observed between 18 May and 2 July 2018 as a part of program no. 15315 (principal investigator: M.M.) using the Advanced Camera for Surveys (ACS) instrument. The F475W and F775W observations consist of 5.04 ks total integration spread across four individual dithered exposures in each band. The F850LP data consist of seven dithered exposures with a 3.96 ks total integration time. These data were reduced using the software package DrizzlePac. Images in each filter were drizzled onto a grid with north up and a pixel size of 0.03 arcsec using the AstroDrizzle routine. This drizzling was done using a Gaussian kernel with a drop size of 0.8 and the individual bands were aligned using the TweakReg routine. The Hubble imaging was then aligned to match the same relative astrometric coordinate frame as the Chandra data by matching the positions of six point sources in the field (X-ray AGN with clear host galaxy counterparts) that are all well detected in both the Hubble imaging and by Chandra (all six with >80 source counts, and four with >200). The residual dispersion in the astrometric alignment between the Hubble and Chandra data is 0.288″.

The extremely long, thin morphology of giant arcs requires non-standard photometric apertures. We generated apertures for the X-ray-emitting arc following the same procedure as in our previous work with strongly lensed arcs[37]. To briefly summarize the process, the aperture was defined by fitting a curve along the ridge line of the arc and then extending orthogonally outward. Photometry of the lensed galaxy was measured in each band of Hubble imaging at an effective aperture radius of 0.35″, and corrected up to a total source flux using the well-calibrated enclosed energy profiles for ACS. The giant arc is quite isolated at Hubble imaging resolutions, resulting in straightforward photometric measurements. In the AB magnitude system the giant arc has $m_{F475W} = 23.66 \pm 0.04$ and $m_{F775W} = 23.41 \pm 0.09$. In F850LP the arc is only marginally detected, with $m_{F850LP} = 23.57 \pm 0.36$. All of these measurements are uncorrected for the lensing magnification, and uncertainties are 1σ confidence intervals.

**Spitzer Space Telescope imaging.** SPT-CLJ2344-4243 was observed with Spitzer/Infrared Array Camera (IRAC)[38] during Cycle 8 as a part of program no. 80012 (principal investigator: M.B.). The observations consisted of 8 × 100 s and 6 × 30 s dithered exposures in IRAC Channel 1 (3.6 μm) and Channel 2 (4.5 μm). These data probe rest-frame wavelengths of ~1–2 μm, providing useful constraints on the total stellar mass (including older stars) in the lensed galaxy. The Channel 1 data are the deeper of the two, with a 10σ point source depth of 20.3 magnitudes and a final effective PSF of 1.66″. We used a reduction of the data produced for previous published work[39], with the same pipeline and procedure as previously described[40]. Spitzer/IRAC photometry was measured using the same apertures that were defined from the Hubble data described in the previous section, convolved with the larger Spitzer/IRAC PSF. The region of sky around the giant arc is relatively isolated in the IRAC imaging, even with the larger PSF. Forced photometry in apertures matched to the Hubble data yields statistical non-detections in both IRAC bands, with flux densities of 5 ± 13 μJy in IRAC Channel 1 (3.6 μm) and 1 ± 23 μJy in IRAC Channel 2 (4.5 μm). From the deeper 3.6 μm data we have placed a 1σ upper limit on the apparent (not corrected for magnification) NIR luminosity (1.5 μm in the rest frame) of $L_{IR,1.5\mu m} < 1.7 \times 10^{10} L_\odot$.

**Gravitational lens modelling.** We computed a strong-lensing model for the foreground galaxy cluster, SPT-CLJ2344-4243, using the public software Lenstool[41]. Lenstool uses a parametric approach, with Markov chain Monte Carlo sampling of the parameter space to identify the best-fit set of parameters and estimate statistical uncertainties. The foreground lens redshift of $z = 0.595$ is estimated from published spectroscopy of 32 cluster members[42]. We describe the cluster mass distribution as a linear combination of several components, each represented by a pseudo-isothermal ellipsoidal mass distribution with seven free parameters: position $x, y$; ellipticity $\theta$, position angle, core and cut radii, and a normalization $\sigma$. As is common practice in computing strong-lens models for clusters of galaxies, the lens plane is modelled with cluster-sized halos, supplemented by less-massive halos placed on individual cluster-member galaxies. For the cluster halos, all the parameters are allowed to vary with the exception of the cut radius that cannot be constrained by the strong-lensing observables alone. The position, ellipticity and position angle of galaxy-scale halos are fixed to their observed properties as measured from the Hubble Space Telescope imaging data using Source Extractor[43]. The core and cut radii and the normalization are scaled to the optical luminosity of each galaxy as measured with Source Extractor. Cluster-member galaxies are selected based on their colour in a colour-magnitude diagram using the red-sequence technique[44].

We identified several sets of multiply-imaged galaxies, whose positions and redshifts are used to constrain the lens model. Preliminary lens models were used in order to identify additional constraints, and lensed galaxies were subsequently spectroscopically confirmed and their redshifts measured with the FIRE spectrograph on the 6.5-m Magellan-I Baade telescope (see section 'Magellan/FIRE infrared spectroscopy'). The best-fit set of parameters is defined as that which minimizes the scatter between the observed and model-predicted images of the strongly lensed galaxies, as measured in the image plane. At the innermost region spanning a few arcseconds around the brightest cluster galaxy, we free the parameters of several halos in order to allow more flexibility in this region that is highly dominated by baryons. This includes the brightest cluster galaxy (BCG), and a component that represents the mass distribution of the core itself, which is allowed to be displaced from the stellar light of the BCG. The parameters of two other halos are set free: a cluster galaxy near the giant arc, and a galaxy-scale halo that accounts for contribution from a foreground galaxy observed at $z = 0.2237$.

We find that the lens plane is adequately modelled by this procedure, with the best fit resulting in an image-plane root-mean-square scatter of 0.37 arcsec. All the sets of multiple images are reproduced by the model. The multiply-imaged lensed galaxies, their positions and redshifts are given in Supplementary Table 2. Supplementary Fig. 4 shows the critical curves of the best-fit model, as well as the positions of the multiple images that were used as constraints, and other background galaxies for which we measured redshifts. We use the final strong-lensing model to generate source-plane reconstructions of the X-ray-emitting giant arc, which are shown in Supplementary Fig. 5.

The reconstruction of the galaxy in the source plane was done by ray-tracing the pixels of the images from the image plane to the source plane, by computing their source-plane position using the lensing equation and the deflection field of the best-fit model. The apparent difference between the two image reconstructions is due to the so-called lensing PSF, caused as the lensing potential distorts the background source. This can be accounted for by forward-modelling the source images, which is beyond the scope of this paper[45,46].

**Nebular emission line diagnostics.** We examined the Balmer emission lines from hydrogen to place a constraint on the amount of dust extinction in the galaxy. The observed ratio of Hα to Hβ emission is consistent with negligible dust extinction, but the uncertainties on the individual line fluxes are large. We place a 2σ (95% confidence) upper limit on the amount of dust extinction at E(B−V) <0.4 magnitudes, using the Calzetti dust attenuation law[47]. The ratio of the strengths of the two [O II] lines in the 3727,3729 doublet constrain[48] the electron density of nebular gas in the lensed galaxy to be approximately 1,000 ± 200 cm$^{-3}$.

Multiple line ratios constrain the relative metal enrichment—measured as the ratio of oxygen to hydrogen atoms in nebular gas. The giant arc is faint, and the large uncertainties on individual line flux measurements in the giant arc limit the precision of any metallicity constraints that we can make, but we did perform the exercise of computing the metallicity, $12 + \log(\text{O/H})$, using several 'strong line' diagnostics. The N2 diagnostic uses the ratio of [N II] to H$\alpha$ emission[49,50] as a proxy for the metallicity, using a locally calibrated scaling relation. This diagnostic estimates $12 + \log(\text{O/H}) = 8.1 \pm 0.4$ for the lensed galaxy. The O3N2 diagnostic[11,50] uses four different lines, as it takes as an input the ratio of two line ratios, [O III]/H$\beta$ divided by [N II]/H$\alpha$; this diagnostic estimates $12 + \log(\text{O/H}) = 8.1 \pm 0.2$. These different diagnostics consistently indicate that the X-ray-emitting arc has a metallicity that is in the range $Z = 0.15$–$0.4\,Z_\odot$ using recent measurements of the solar $12 + \log(\text{O/H})$ metallicity[51].

**Star-formation rate and stellar mass estimates.** We estimated the SFR of the lensed X-ray arc from both the measured H$\alpha$ emission line and the rest-frame UV stellar continuum emission. The SFR was estimated by computing the observed H$\alpha$ luminosity of the giant arc and then following the standard prescription[52] for Case B recombination, and including an aperture correction for the fraction of the giant arc that fell within the FIRE slit. The resulting apparent H$\alpha$ SFR is $20 \pm 4\,M_\odot\,\text{yr}^{-1}$. We then computed the true, intrinsic SFR by correcting for the average magnification affecting the portion of the arc that fell within the FIRE slits—a factor of $65 \pm 20$—resulting in $\text{SFR}_{\text{H}\alpha} = 0.4 \pm 0.1\,M_\odot\,\text{yr}^{-1}$, where the reported uncertainty is the 1$\sigma$ confidence interval. This measurement represents a minimum SFR, assuming no dust extinction and associated obscured star formation. We used the upper limit on the extinction from the hydrogen Balmer line ratio to compute an upper limit on the SFR of $1.2\,M_\odot\,\text{yr}^{-1}$.

We computed a second SFR estimate based on the observed UV stellar continuum emission[52] using the Hubble F475W imaging, which samples a rest-frame wavelength of $\sim 1{,}900$ Å for the giant arc. Using the redshift and magnification of the giant arc we first computed the absolute UV magnitude at the giant arc at rest-frame 1,900 Å to be $M_{\text{UV}} = -16.3 \pm 0.4$. The scaling between UV luminosity and the SFR is sensitive to the properties of the underlying stellar population, such that the inferred SFR from a single UV luminosity changes by 42% for a 100 Myr versus 10 Myr old stellar population[53]. If the X-ray emission that we observe is associated with HMBXs, as we argue below, then the observed emission from the giant arc in SPT-CLJ2344-4243 is likely to be dominated by a young ($<30$ Myr) population. We can infer that the stellar population of the X-ray-emitting arc is young, but we lack a precise measurement of its age. We therefore use the SFR scaling estimator calibrated for a very young (10 Myr) population, but we note that the $L_{\text{UV}}$-to-SFR scaling factor varies systematically with stellar population age, increasing as age decreases. To reflect this fact, we include an additional 40% systematic uncertainty in the stellar population age that reflects the range of $L_{\text{UV}}$-to-SFR conversions between 10 and 100 Myr old stellar populations. Assuming a young stellar population age results in an apparent UV continuum SFR estimate of $19 \pm 8\,M_\odot\,\text{yr}^{-1}$, uncorrected for the lensing magnification. Applying a magnification correction using the light-weighted average magnification—a factor of $60 \pm 20$—for the photometric aperture that we used to measure the F475W flux, we recover an intrinsic $\text{SFR}_{\text{UV}} = 0.2 \pm 0.1\,M_\odot\,\text{yr}^{-1}$. If we apply the upper limit on the rest-frame extinction from above, then the resulting constraint is $\text{SFR}_{\text{UV}} < 1.0\,M_\odot\,\text{yr}^{-1}$. Ideally, we would also incorporate a rest-frame infrared (IR) measurement of the SFR to directly constrain the amount of obscured star formation, but because we have no data on the giant arc at wavelengths longer than 4.5 $\mu$m (1.8 $\mu$m in the rest frame), we must rely solely on extinction-corrected optical and UV estimates of the total SFR. We are encouraged that the SFRs derived from the UV continuum and H$\alpha$ emission line flux are in excellent agreement, despite being subject to very different extinction correction factors.

Throughout this paper we use SFR estimators for both H$\alpha$ and the UV luminosity that are calibrated assuming a Salpeter[54] initial mass function (IMF), and for consistency, where necessary, we have converted all literature observations and models against which we compare our measurements to a common Salpeter IMF. If we were to assume an IMF more in line with the currently preferred form[55], then the resulting H$\alpha$- and UV-based SFR estimates would be 68% and 63% smaller, respectively. Because the UV-based SFR estimate is subject to substantial systematic uncertainty due to assumptions about the age of the stellar population, we elect to use $\text{SFR}_{\text{H}\alpha}$ as the best-available measurement. More precise constraints on the SFR of the X-ray-emitting arc would require additional data, such as deep multi-band NIR through optical imaging and deeper spectroscopy that is likely only feasible with future facilities such as the James Webb Space Telescope or 30-meter class ground-based telescopes.

The ideal method for constraining the stellar mass of the giant arc would have been to perform a multi-band fit of a suite of galaxy templates to the spectral energy distribution (SED) of the arc. However, with only two robust detections in adjacent Hubble bands sampling the rest-frame UV and upper limits on the rest-frame NIR from Spitzer, the SED of the galaxy cannot usefully constrain the shape of the SED. Given the dearth of constraining information about the stellar population in the lensed galaxy, we have placed a limit on the stellar mass by applying empirically calibrated stellar mass-to-light (M/L) ratio fitting formulae[56] that use the rest-frame NIR luminosity and measured colour. We used the upper limit of the luminosity at 3.6 $\mu$m (rest frame 1.5 $\mu$m) and the observed UV colour from the Hubble imaging that were measured as described above, and applied the published M/L relation, corrected to a Salpeter IMF to be consistent with our SFR estimates, and inferred an upper limit on the apparent (not corrected for magnification) stellar mass of $6.3 \times 10^9\,M_\odot$. Using the magnification estimate computed from the strong-lensing model described above for the photometric aperture defined in the Hubble imaging, the upper limit on the intrinsic (true) stellar mass of the lensed galaxy is $1 \times 10^8\,M_\odot$. Given the dearth of data available, we consider this to be only a very crude constraint on the stellar mass of the lensed galaxy, and note that it is subject to substantial systematic uncertainties. Our primary objective here is merely to get a rough handle on the stellar mass in the lensed galaxy so that we can examine it in the context of other samples spanning several decades in stellar mass. In that context, an order of magnitude stellar mass limit is sufficient to the task.

**The origin of the observed X-ray emission.** We have used multi-wavelength follow-up data to determine whether the observed X-ray emission is associated with star formation or AGN activity. The best constraints come from the ratios of strong rest-frame optical emission lines that we plot on the classic Baldwin, Phillips and Terlevich (BPT) diagram in Supplementary Fig. 6, which identifies the dominant sources of ionizing radiation in galaxies[10,11]. In the BPT diagram we see that the X-ray-emitting lensed galaxy in SPT-CLJ2344-4243 lies in the region occupied by $z \sim 0$ star-forming galaxies in the Sloan Digital Sky Survey (SDSS)[57], as well as dwarf starburst galaxies in the local Universe that have measured X-ray emission from star formation (Supplementary Fig. 6)[14–16,21]. In contrast, known low-luminosity AGNs in the local Universe with X-ray luminosities similar to our lensed galaxy[58] still sit squarely within the AGN region of the BPT diagram. We also note that the rest-frame optical lines used to place the X-ray arc in the BPT diagram resulted from observations in which the vast majority (~90%) of the integration time had the FIRE slit positioned so as to fall directly on top of the X-ray emission. These spectra would, therefore, have been sensitive to even AGN-like BPT line ratios that could be localized to the portion of the galaxy that is responsible for emitting in the X-ray. In AGN contamination models based on SDSS galaxies, a galaxy with BPT diagnostic ratios similar to the X-ray-emitting giant arc ([O III]/H$\beta = 0.6$, [N II]/H$\alpha = 0.03$) has a minimal chance (consistent with ~0–1%) of hosting an AGN[59]. This is corroborated by detailed studies that disentangle the X-ray emission from an AGN and star formation in the cores of local galaxies[60], in which only one out of 51 galaxies studied host an AGN while exhibiting a [N II]/H$\alpha < 0.1$, and even that galaxy has BPT ratios that differ substantially (lower in [O III]/H$\beta$ by a factor of two and in [N II]/H$\alpha$ by a factor of three) from the X-ray-emitting giant arc. We do not have any data constraining the properties of the lensed galaxy at wavelengths longer than ~2 $\mu$m, and so we cannot test for the presence of an IR excess, but such a signature would also be quite surprising given the low dust content and stellar mass of the lensed galaxy.

Examining the available X-ray data, we compared the observed hardness ratio of the lensed source to samples of distant galaxies and AGN with X-ray detections. The limited available counts results in a large uncertainty in the hardness, $0.2 \pm 0.3$, which, but this value is broadly consistent with measured hardness ratios of the $z \gtrsim 1$ star-forming galaxies in X-ray deep fields, which tend to have positive hardness ratios[20,33]. Distant AGN can have a broad range of hardness ratios, depending on the column density of absorbing neutral hydrogen ($N_\text{H}$), but less obscured AGN ($N_\text{H} \lesssim$ a few $\times 10^{23}\,\text{cm}^{-2}$) tend to have negative hardness ratios[61].

We also look to the source-plane reconstruction of all of the observed emission from the lensed galaxy for evidence of its origin. Supplementary Fig. 5 reveals that the giant arc is intrinsically a small ($r \lesssim 1$ kpc) blue galaxy with UV emission coming from two spatially resolved star-forming regions separated by ~500 pc in projection. The FIRE spectroscopy includes slits that targeted each of the two star-forming clumps, confirming that they share a common redshift, being identical to within the measurement uncertainties (0.00012 in redshift, 30 km s$^{-1}$ in velocity). The Chandra PSF is sufficiently small to localize the X-ray emission to one of the two UV-bright clumps in the lensed galaxy, specifically, the northwestern most (upper right clump in the source-plane images; Supplementary Fig. 5) of the two UV-bright star-forming regions. As discussed above, the point-like nature of the X-ray emission implies that the emitting source(s) are confined to a region with diameter $\lesssim 400$ pc. This size is comparable to large star-forming regions, so that the unresolved X-ray detection of this lensed galaxy is consistent with what a population of HMXBs formed in a large star-forming region (or complex of smaller star-forming regions) within the lensed galaxy.

Finally, we also consider the possibility that the X-ray emission results from a chance projection of a background X-ray source onto the location of the optical giant arc. The primary evidence against this possibility is the precise alignment between the two X-ray sources and the two strongly lensed optical images of the galaxy. The strong-lensing configuration here is that of a merging image pair, and in such a configuration there is a parity flip between the two images. The optical and X-ray images of the lensed galaxy exhibit this parity flip exactly as predicted for strong lensing, while the scenario in which these images are random projections is exceedingly unlikely, as it would require the random coincidence in location of two X-ray sources with the two ends of the optical giant arc. The surface density of X-ray sources in deep fields is approximately 1 every 1,000 square arcseconds[20], and so the odds of two such sources randomly falling on two specific locations in the sky, localized to the precision of our data (~0.3″), is approximately 1 in $10^7$. The odds of this happening become vanishingly small when we also consider that the two X-ray images have the same observed X-ray flux (to within the uncertainties), as would be expected for two similarly magnified images of the same galaxy.

Given all of the available evidence, we have concluded that the X-ray emission detected from the giant arc in SPT-CLJ2344-4243 is generated predominantly by stellar-mass compact object systems associated with recent star formation. The available data cannot absolutely rule out the presence of a very faint/obscured AGN, but we argue that this interpretation for the observed X-ray emission is much less likely than HMXBs. We come to this conclusion based on the weight of the evidence, including the BPT lines, the nature and morphology of the galaxy (consistent with a low-mass dwarf starburst), and the X-ray emission being clearly associated with a UV-bright star-forming region that is spatially resolved in the Hubble imaging.

Observations of local dwarf starburst galaxies have demonstrated that it is common for these objects to have X-ray emission resulting from HMXBs and ultra-luminous X-ray sources associated with young stellar populations[15,25,27]. This picture is further supported by our nearest example of a starburst galaxy, M82, in which the individual HMXB sources have been spatially resolved and well studied[62]. There have been two cases reporting potential low-luminosity X-ray AGN in dwarf starburst galaxies[63,64] that fall into the star-forming region of the BPT diagram as the giant arc analysed here. However, a deeper follow-up X-ray study of one of these galaxies revealed that the dominant source of X-ray emission is likely to be HMXBs associated with star-forming regions ($L_X \sim 10^{39}$–$10^{40}$ erg s$^{-1}$), with any AGN that is present being sub-Eddington and much fainter[65] ($L_X \sim 10^{38}$ erg s$^{-1}$). We conclude that the weight of the evidence overwhelmingly points toward star formation as the source of the observed X-ray emission in the giant arc.

## Data availability

The data that support the plots within this paper and other findings of this study are available from the corresponding author upon reasonable request. This paper makes use of Chandra data from observation IDs 13401, 16135, 16545, 19581, 19582, 19583, 20630, 20631, 20634, 20635, 20636 and 20797. All raw Chandra data are available for download from the Chandra X-ray Center (https://cda.harvard.edu/chaser/). The Hubble data used in this work is available at the Mikulski Archive for Space Telescopes (MAST; https://archive.stsci.edu) under proposal ID 15315. The full raw and reduced FIRE spectroscopy used in this work is freely available upon request. The reduced spectrum is publicly available for download at the Harvard Dataverse (https://dataverse.harvard.edu/dataset.xhtml?persistentId=doi:10.7910/DVN/JCFRLB).

## Code availability

The data reduction pipelines used in this work are all publicly available. Chandra data were reduced using the CIAO package (http://cxc.harvard.edu/ciao/), Hubble data were reduced using Drizzlepac (http://drizzlepac.stsci.edu/), and FIRE data were reduced using the FIREHOSE package (http://web.mit.edu/rsimcoe/www/FIRE/ob_data.htm). The modelling of the foreground galaxy cluster potential was done using the publicly available Lenstool code (https://projets.lam.fr/projects/lenstool/wiki). Analysis of the FIRE spectra was performed using the IDL Astronomy User's Library (https://idlastro.gsfc.nasa.gov/).

## References

1. Coe, D. et al. CLASH: three strongly lensed images of a candidate $z \sim 11$ galaxy. *Astrophys. J.* **762**, 32 (2013).
2. Vieira, J. D. et al. Dusty starburst galaxies in the early Universe as revealed by gravitational lensing. *Nature* **495**, 344–347 (2013).
3. Ishigaki, M. et al. Hubble Frontier Fields first complete cluster data: faint galaxies at $z \sim 5$–10 for UV luminosity functions and cosmic reionization. *Astrophys. J.* **799**, 12 (2015).
4. Livermore, R. C., Finkelstein, S. L. & Lotz, J. M. Directly observing the galaxies likely responsible for reionization. *Astrophys. J.* **835**, 113 (2017).
5. Laporte, N. et al. Dust in the reionization era: ALMA observations of a $z = 8.38$ gravitationally lensed galaxy. *Astrophys. J. Lett.* **837**, L21 (2017).
6. Lotz, J. M. et al. The Frontier Fields: survey design and initial results. *Astrophys. J.* **837**, 97 (2017).
7. Planck Collaboration Planck 2018 results. VI. Cosmological parameters. Preprint at https://arxiv.org/abs/1807.06209 (2018).
8. Carlstrom, J. E. et al. The 10 meter South Pole Telescope. *Publ. Astron. Soc. Pac.* **123**, 568–581 (2011).
9. McDonald, M. et al. A massive, cooling-flow-induced starburst in the core of a luminous cluster of galaxies. *Nature* **488**, 349–352 (2012).
10. Baldwin, A., Phillips, M. M. & Terlevich, R. Classification parameters for the emission-line spectra of extragalactic objects. *Publ. Astron. Soc. Pac.* **93**, 5–19 (1981).
11. Kewley, L. J. et al. The cosmic BPT diagram: confronting theory with observations. *Astrophys. J. Lett.* **774**, L10 (2013).
12. Fragos, T. et al. X-ray binary evolution across cosmic time. *Astrophys. J.* **764**, 41 (2013).
13. Colbert, E. J. M., Heckman, T. M., Ptak, A. F., Strickland, D. K. & Weaver, K. A. Old and young X-ray point source populations in nearby galaxies. *Astrophys. J.* **602**, 231–248 (2004).
14. Basu-Zych, A. R. et al. Evidence for elevated X-ray emission in local Lyman break galaxy analogs. *Astrophys. J.* **774**, 152 (2013).
15. Brorby, M., Kaaret, P., Prestwich, A. & Mirabel, I. F. Enhanced X-ray emission from Lyman break analogues and a possible $L_X$-SFR-metallicity plane. *Mon. Not. R. Astron. Soc.* **457**, 4081–4088 (2016).
16. Kaaret, P., Brorby, M., Casella, L. & Prestwich, A. H. Resolving the X-ray emission from the Lyman-continuum emitting galaxy Tol 1247–232. *Mon. Not. R. Astron. Soc.* **471**, 4234–4238 (2017).
17. Svoboda, J., Douna, V., Orlitová, I. & Ehle, M. Green Peas in X-rays. *Astrophys. J.* **880**, 144–159 (2019).
18. Lehmer, B. D. et al. The evolution of normal galaxy X-ray emission through cosmic history: constraints from the 6 ms Chandra Deep Field-South. *Astrophys. J.* **825**, 7 (2016).
19. Xue, Y. Q. et al. The 2 ms Chandra Deep Field-North Survey and the 250 ks extended Chandra Deep Field-South Survey: improved point-source catalogs. *Astrophys. J. Suppl.* **224**, 15 (2016).
20. Luo, B. et al. The Chandra Deep Field-South Survey: 7 ms source catalogs. *Astrophys. J. Suppl.* **228**, 2 (2017).
21. Basu-Zych, A. R. et al. The X-ray star formation story as told by Lyman break galaxies in the 4 ms CDF-S. *Astrophys. J.* **762**, 45 (2013).
22. Aird, J., Coil, A. L. & Georgakakis, A. X-rays across the galaxy population—I. Tracing the main sequence of star formation. *Mon. Not. R. Astron. Soc.* **465**, 3390–3415 (2017).
23. Mineo, S., Gilfanov, M. & Sunyaev, R. X-ray emission from star-forming galaxies—I. High-mass X-ray binaries. *Mon. Not. R. Astron. Soc.* **419**, 2095–2115 (2012).
24. Ranalli, P., Comastri, A. & Setti, G. The 2–10 keV luminosity as a star formation rate indicator. *Astron. Astrophys.* **399**, 39–50 (2003).
25. Prestwich, A. H. et al. Ultra-luminous X-ray sources in HARO II and the role of X-ray binaries in feedback in Lyα emitting galaxies. *Astrophys. J.* **812**, 166 (2015).
26. Mineo, S., Gilfanov, M., Lehmer, B. D., Morrison, G. E. & Sunyaev, R. X-ray emission from star-forming galaxies—III. Calibration of the $L_X$-SFR relation up to redshift $z \sim 1.3$. *Mon. Not. R. Astron. Soc.* **437**, 1698–1707 (2014).
27. Basu-Zych, A. R. et al. Exploring the overabundance of ULXs in metal- and dust-poor local Lyman break analogs. *Astrophys. J.* **818**, 140 (2016).
28. Mesinger, A., Ferrara, A. & Spiegel, D. S. Signatures of X-rays in the early Universe. *Mon. Not. R. Astron. Soc.* **431**, 621–637 (2013).
29. Fialkov, A., Barkana, R. & Visbal, E. The observable signature of late heating of the Universe during cosmic reionization. *Nature* **506**, 197–199 (2014).
30. Das, A., Mesinger, A., Pallottini, A., Ferrara, A. & Wise, J. H. High-mass X-ray binaries and the cosmic 21-cm signal: impact of host galaxy absorption. *Mon. Not. R. Astron. Soc.* **469**, 1166–1174 (2017).
31. McDonald, M. et al. The growth of cool cores and evolution of cooling properties in a sample of 83 galaxy clusters at $0.3 < z < 1.2$ selected from the SPT-SZ Survey. *Astrophys. J.* **774**, 23 (2013).
32. McDonald, M. et al. Deep Chandra, HST-COS, and Megacam observations of the Phoenix cluster: extreme star formation and AGN feedback on hundred kiloparsec scales. *Astrophys. J.* **811**, 111 (2015).
33. Ranalli, P. et al. X-ray properties of radio-selected star forming galaxies in the Chandra-COSMOS survey. *Astron. Astrophys.* **542**, A16 (2012).
34. Simcoe, R. A. et al. FIRE: a near-infrared cross-dispersed echellette spectrometer for the Magellan telescopes. *Proc. SPIE* **7014**, 70140U (2008).
35. Vacca, W. D., Cushing, M. C. & Rayner, J. T. A method of correcting near-infrared spectra for telluric absorption. *Publ. Astron. Soc. Pac.* **115**, 389–409 (2003).
36. Cushing, M. C., Vacca, W. D. & Rayner, J. T. Spextool: a spectral extraction package for SpeX, a 0.8-5.5 micron cross-dispersed spectrograph. *Publ. Astron. Soc. Pac.* **116**, 362–376 (2004).
37. Bayliss, M. B. Broadband photometry of 105 giant arcs: redshift constraints and implications for giant arc statistics. *Astrophys. J.* **744**, 156 (2012).
38. Fazio, G. G. et al. The Infrared Array Camera (IRAC) for the Spitzer Space Telescope. *Astrophys. J. Suppl.* **154**, 10–17 (2004).
39. Bleem, L. E. et al. Galaxy clusters discovered via the Sunyaev-Zel'dovich effect in the 2500-square-degree SPT-SZ survey. *Astrophys. J. Suppl.* **216**, 27 (2015).
40. Ashby, M. L. N. et al. The Spitzer Deep, Wide-Field Survey. *Astrophys. J.* **701**, 428–453 (2009).
41. Jullo, E. et al. A Bayesian approach to strong lensing modelling of galaxy clusters. *New J. Phys.* **9**, 447 (2007).
42. Bayliss, M. B. et al. SPT-GMOS: a Gemini/GMOS-South spectroscopic survey of galaxy clusters in the SPT-SZ survey. *Astrophys. J. Suppl.* **227**, 3 (2016).

43. Bertin, E. & Arnouts, S. SExtractor: software for source extraction. *Astron. Astrophys. Suppl.* **117**, 393–404 (1996).
44. Gladders, M. D. & Yee, H. K. C. A new method for galaxy cluster detection. I. The algorithm. *Astron. J.* **120**, 2148–2162 (2000).
45. Johnson, T. L. et al. Star formation at $z = 2.481$ in the lensed galaxy SDSS J1110=6459. I. Lens modeling and source reconstruction. *Astrophys. J.* **843**, 78 (2017).
46. Sharma, S. et al. High-resolution spatial analysis of a $z \sim 2$ lensed galaxy using adaptive coadded source-plane reconstruction. *Mon. Not. R. Astron. Soc.* **481**, 1427–1440 (2018).
47. Calzetti, D. The dust opacity of star-forming galaxies. *Publ. Astron. Soc. Pac.* **113**, 1449–1485 (2001).
48. Osterbrock, D. E. *Astrophysics of Gaseous Nebulae and Active Galactic Nuclei* (University Science Books, 1989).
49. Denicoló, G., Terlevich, R. & Terlevich, E. New light on the search for low-metallicity galaxies—I. The $N_2$ calibrator. *Mon. Not. R. Astron. Soc.* **330**, 69–74 (2002).
50. Pettini, M. & Pagel, B. E. J. [O III]/[N II] as an abundance indicator at high redshift. *Mon. Not. R. Astron. Soc.* **348**, L59–L63 (2004).
51. Asplund, M., Grevesse, N., Sauval, A. J. & Scott, P. The chemical composition of the Sun. *Annu. Rev. Astron. Astrophys.* **47**, 481–522 (2009).
52. Kennicutt, R. C. & Evans, N. J. Star formation in the Milky Way and nearby galaxies. *Annu. Rev. Astron. Astrophys.* **50**, 531–608 (2012).
53. Leitherer, C. et al. Starburst99: synthesis models for galaxies with active star formation. *Astrophys. J. Suppl.* **123**, 3–40 (1999).
54. Salpeter, E. E. The luminosity function and stellar evolution. *Astrophys. J.* **121**, 161 (1955).
55. Kroupa, P. On the variation of the initial mass function. *Mon. Not. R. Astron. Soc.* **322**, 231–246 (2001).
56. Bell, E. F. & de Jong, R. S. Stellar mass-to-light ratios and the Tully-Fisher relation. *Astrophys. J.* **550**, 212–229 (2001).
57. Thomas, D. et al. Stellar velocity dispersions and emission line properties of SDSS-III/BOSS galaxies. *Mon. Not. R. Astron. Soc.* **431**, 1383–1397 (2013).
58. Terashima, Y., Iyomoto, N., Ho, L. C. & Ptak, A. F. X-Ray properties of LINERs and low-luminosity Seyfert galaxies observed with ASCA. I. Observations and results. *Astrophys. J. Suppl.* **139**, 1–36 (2002).
59. Jones, M. L. et al. The intrinsic Eddington ratio distribution of active galactic nuclei in star-forming galaxies from the Sloan Digital Sky Survey. *Astrophys. J.* **826**, 12 (2016).
60. She, R., Ho, L. C. & Feng, H. Chandra survey of nearby galaxies: a significant population of candidate central black holes in late-type galaxies. *Astrophys. J.* **842**, 131 (2017).
61. Wang, J. X., Malhotra, S., Rhoads, J. E. & Norman, C. A. Identifying high-redshift active galactic nuclei using X-ray hardness. *Astrophys. J.* **612**, L109–L112 (2004).
62. Kong, A. K. H., Yang, Y. J., Hsieh, P.-Y., Mak, D. S. Y. & Pun, C. S. J. The ultraluminous X-ray sources near the center of M82. *Astrophys. J.* **671**, 349–357 (2007).
63. Reines, A. E., Sivakoff, G. R., Johnson, K. E. & Brogan, C. L. An actively accreting massive black hole in the dwarf starburst galaxy Henize2–10. *Nature* **470**, 66–68 (2011).
64. Reines, A. E. et al. A candidate massive black hole in the low-metallicity dwarf galaxy pair Mrk 709. *Astrophys. J. Lett.* **787**, L30 (2014).
65. Reines, A. E. et al. Deep Chandra observations of the compact starburst galaxy Henize 2–10: X-rays from the massive black hole. *Astrophys. J. Lett.* **830**, L35 (2016).

### Acknowledgements
Support for this work was provided by NASA through Chandra award number GO7-18124, issued by the Chandra X-ray Observatory Center, which is operated by the Smithsonian Astrophysical Observatory for and on behalf of the National Aeronautics Space Administration under contract NAS8-03060. Additional support was provided by NASA through the Space Telescope Science Institute (HST-GO-15315), which is operated by the Association of Universities for Research in Astronomy, Incorporated, under NASA contract NAS5-26555.

### Author contributions
M.B.B. performed the analysis of the FIRE spectroscopy, Chandra, Hubble and Spitzer data, and wrote the article text with input and contributions from all authors. M.M. acquired the Chandra and Hubble data. M.B.B. and M.M. reduced the Chandra X-ray data. M.B.B. reduced the FIRE NIR spectroscopy, which was obtained from observations performed by M.B.B. and M.D.G. M.B. acquired the Spitzer data. M.F. reduced the Hubble data. K.S. computed the strong lensing model of the foreground cluster lens and produced the reconstruction of the galaxy in the source plane. The authors are ordered in two alphabetical tiers after M.F.

### Competing interests
The authors declare no competing interests.

### Additional information
**Supplementary information** is available for this paper at https://doi.org/10.1038/s41550-019-0888-7.

**Correspondence and requests for materials** should be addressed to M.B.B.

**Peer review information** *Nature Astronomy* thanks Jean-Paul Kneib and the other, anonymous, reviewer(s) for their contribution to the peer review of this work.

**Reprints and permissions information** is available at www.nature.com/reprints.

**Publisher's note** Springer Nature remains neutral with regard to jurisdictional claims in published maps and institutional affiliations.

# An X-ray Detection of Star Formation In a Highly Magnified Giant Arc

M. B. Bayliss[1,2,*], M. McDonald[1], K. Sharon[3], M. D. Gladders[4,5], M. Florian[6], J. Chisholm[7], H. Dahle[8], G. Mahler[3], R. Paterno-Mahler[9], J. R. Rigby[6], E. Rivera-Thorsen[8], K. E. Whitaker[10,11,12], S. Allen[13,14,15], B. A. Benson[4,5,16], L. E. Bleem[17], M. Brodwin[18], R. E. A. Canning[13,14], I. Chiu[19], J. Hlavacek-Larrondo[20], G. Khullar[4,5], C. Reichardt[21], J. D. Vieira[22]

[1]*Kavli Institute for Astrophysics & Space Research, Massachusetts Institute of Technology, 77 Massachusetts Avenue, Cambridge, MA 02139, USA*
[2]*Department of Physics, University of Cincinnati, Cincinnati, OH 45221, USA*
[3]*Department of Astronomy, University of Michigan, 1130 South University Ave., Ann Arbor, MI, 48109, USA*
[4]*Department of Astronomy and Astrophysics, University of Chicago, 5640 South Ellis Avenue, Chicago, IL 60637, USA*
[5]*Kavli Institute for Cosmological Physics, University of Chicago, 5640 South Ellis Avenue, Chicago, IL 60637, USA*
[6]*Observational Cosmology Lab, Goddard Space Flight Center, 8800 Greenbelt Road, Greenbelt, MD 20771, USA*
[7]*Department of Astronomy and Astrophysics, University of California, Santa Cruz, CA 95064, USA*
[8]*Institute of Theoretical Astrophysics, University of Oslo, PO Box 1029, Blindern, 0315, Oslo, Norway*
[9]*Department of Physics and Astronomy, University of California, Irvine, 4129 Frederick Reines Hall, Irvine, CA 92697, USA*
[10]*Department of Astronomy, University of Massachusetts, Amherst, MA 01003, USA*
[11]*Department of Physics, University of Connecticut, 2152 Hillside Rd. Unit 3046, Storrs, CT 06269, USA*
[12]*Cosmic Dawn Center (DAWN), Vibenshuset, Lyngbyvej 2, 2100 Copenhagen, Denmark*
[13]*Kavli Institute for Particle Astrophysics and Cosmology (KIPAC), Stanford University, 452 Lomita Mall, Stanford, CA 94305-4085, USA*
[14]*Department of Physics, Stanford University, 452 Lomita Mall, Stanford, CA 94305-4085, USA*
[15]*SLAC National Accelerator Laboratory, 2575 Sand Hill Road, Menlo Park, CA 94025, USA*
[16]*Fermi National Accelerator Laboratory, Batavia, IL 60510-0500, USA*
[17]*Argonne National Laboratory, High-Energy Physics Division, 9700 S. Cass Avenue, Argonne,*

IL 60439, USA

[18] *Department of Physics and Astronomy, University of Missouri, 5110 Rockhill Road, Kansas City, MO 64110, USA*

[19] *Academia Sinica Institute of Astronomy and Astrophysics, 11F of AS/NTU Astronomy-Mathematics Building, No.1, Sec. 4, Roosevelt Road, Taipei 10617, Taiwan*

[20] *Department of Physics, University of Montreal, Montreal, QC H3C 3J7, Canada*

[21] *School of Physics, University of Melbourne, Parkville, VIC 3010, Australia*

[22] *Department of Astronomy and Department of Physics, University of Illinois, 1002 West Green Street, Urbana, IL 61801, USA*

Supplementary Information for "An X-ray Detection of Star Formation In a Highly Magnified Giant Arc":

**Supplementary Table 1** | FIRE Emission Line Fluxes

| Ion | Rest-frame Wavelength (Å) | $f_\lambda$ ($\times 10^{-18}$ erg cm$^{-2}$ s$^{-1}$) | 1 $\sigma$ error on $f_\lambda$ ($\times 10^{-18}$ erg cm$^{-2}$ s$^{-1}$) |
|---|---|---|---|
| **[O II]** | 3727.10 | 6.1 | 1.6 |
| **[O II]** | 3729.86 | 6.3 | 1.7 |
| **H$\beta$** | 4862.70 | 5.1 | 1.0 |
| **[O III]** | 4960.29 | 6.5 | 1.1 |
| **[O III]** | 5008.24 | 13.9 | 1.8 |
| **H$\alpha$** | 6564.63 | 11.1 | 1.5 |
| **[N II]** | 6585.42 | 0.4 | 0.3 |

## Supplementary Table 2 | List of Lensing Constraints

| ID | R.A. J2000 | Decl. J2000 | $z_{\rm spec}$ | $z_{\rm spec}$ Source | Notes |
|---|---|---|---|---|---|
| 1.1 | 356.181065 | -42.719537 | 1.1485 | FIRE | Radial arc |
| 1.2 | 356.180015 | -42.719137 | 1.1485 | FIRE | Radial arc |
| 1.3 | 356.194399 | -42.721999 | | | |
| 2.1 | 356.193637 | -42.718398 | 1.5244 | FIRE | X-ray source |
| 2.2 | 356.192454 | -42.716205 | | FIRE | |
| 2.3 | 356.179454 | -42.713201 | | | Predicted counter image |
| 2.4 | 356.175115 | -42.729047 | | | Predicted counter image |
| 3.1 | 356.182510 | -42.716901 | 1.5130 | FIRE | Radial arc |
| 3.2 | 356.182793 | -42.718304 | | | Radial arc |
| 3.3 | 356.178356 | -42.732671 | | | |
| 4.1 | 356.188161 | -42.718569 | 3.1854 | FIRE | |
| 4.2 | 356.187273 | -42.718805 | | | |
| 5.1 | 356.165815 | -42.719573 | 1.9760 | FIRE | |
| 5.2 | 356.165753 | -42.719721 | | | |
| 5.3 | 356.165866 | -42.720521 | | | |
| 5.4 | 356.190844 | -42.724246 | | | |
| 5a.3 | 356.165860 | -42.720559 | | | |
| 5a.4 | 356.190861 | -42.724208 | | | |
| 5b.3 | 356.165787 | -42.720653 | | | |
| 5b.4 | 356.190914 | -42.724083 | | | |
| 6.1 | 356.164709 | -42.720192 | — | | |
| 6.2 | 356.164672 | -42.720768 | | | |
| 8.1 | 356.189149 | -42.719243 | 1.4278 | FIRE | |
| 8.2 | 356.169401 | -42.726908 | | | |
| 8.3 | 356.184586 | -42.719816 | | | |
| 8a.1 | 356.189016 | -42.719262 | | | |
| 8a.2 | 356.169315 | -42.726879 | | | |
| 9.1 | 356.181472 | -42.714693 | 1.5080 | FIRE | |
| 9.3 | 356.177247 | -42.731010 | | | |
| 10.1 | 356.194985 | -42.720406 | 2.9181 | FIRE | |
| 10.2 | 356.165524 | -42.728406 | | | |
| 10a.1 | 356.165609 | -42.728576 | | | |
| 10a.2 | 356.194923 | -42.720238 | | | |
| A | 356.169416 | -42.720220 | 1.2159 | FIRE | Single image |
| B | 356.169143 | -42.722700 | 1.2150 | FIRE | Single image |
| C | 356.191627 | -42.715342 | — | | |
| D | 356.194568 | -42.712562 | — | FIRE | Observed, no spec-z |
| E | 356.169583 | -42.716381 | — | | |
| F | 356.187354 | -42.727401 | 0.6500 | Bayliss+2016 | |
| G | 356.171620 | -42.725820 | 0.2237 | Bayliss+2016 | |
| Cluster | 356.18314 | -42.720148 | 0.596 | Bayliss+2016 | |

The Right Ascension (R.A.) and Declination (Decl.) for all galaxies used as inputs for the strong lensing model described in Methods Section 5. All lensed background redshifts were obtained with the FIRE spectrograph on Magellan as described in the Methods Section 2. Cluster and foreground galaxy redshifts are taken from the literature[1].

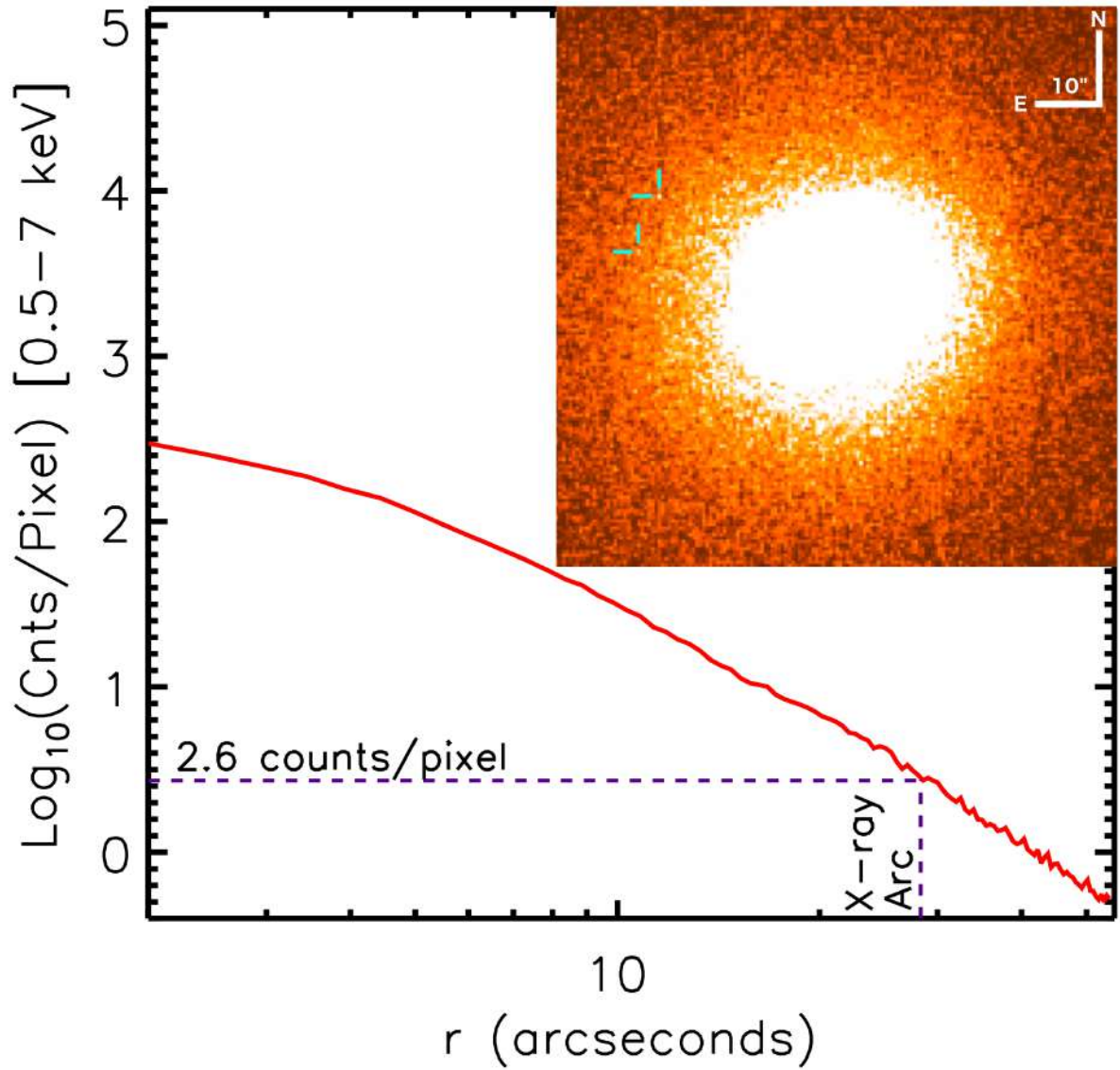

**Supplementary Figure 1 | X-ray Surface Brightness Profile.** The surface brightness profile of the foreground lensing galaxy cluster, SPT-CLJ2344-4243, is well measured out to a radius of more than 60 arcseconds, and in the region where the giant arc is located the cluster emission is spatially smooth, with 2.6 counts per *Chandra* pixel. The reduced 0.5-7 keV *Chandra* image used to measure the cluster surface brightness profile is inset in the upper right. The emission from the giant arc is visible even in the raw frame (indicated by the cyan hash marks).

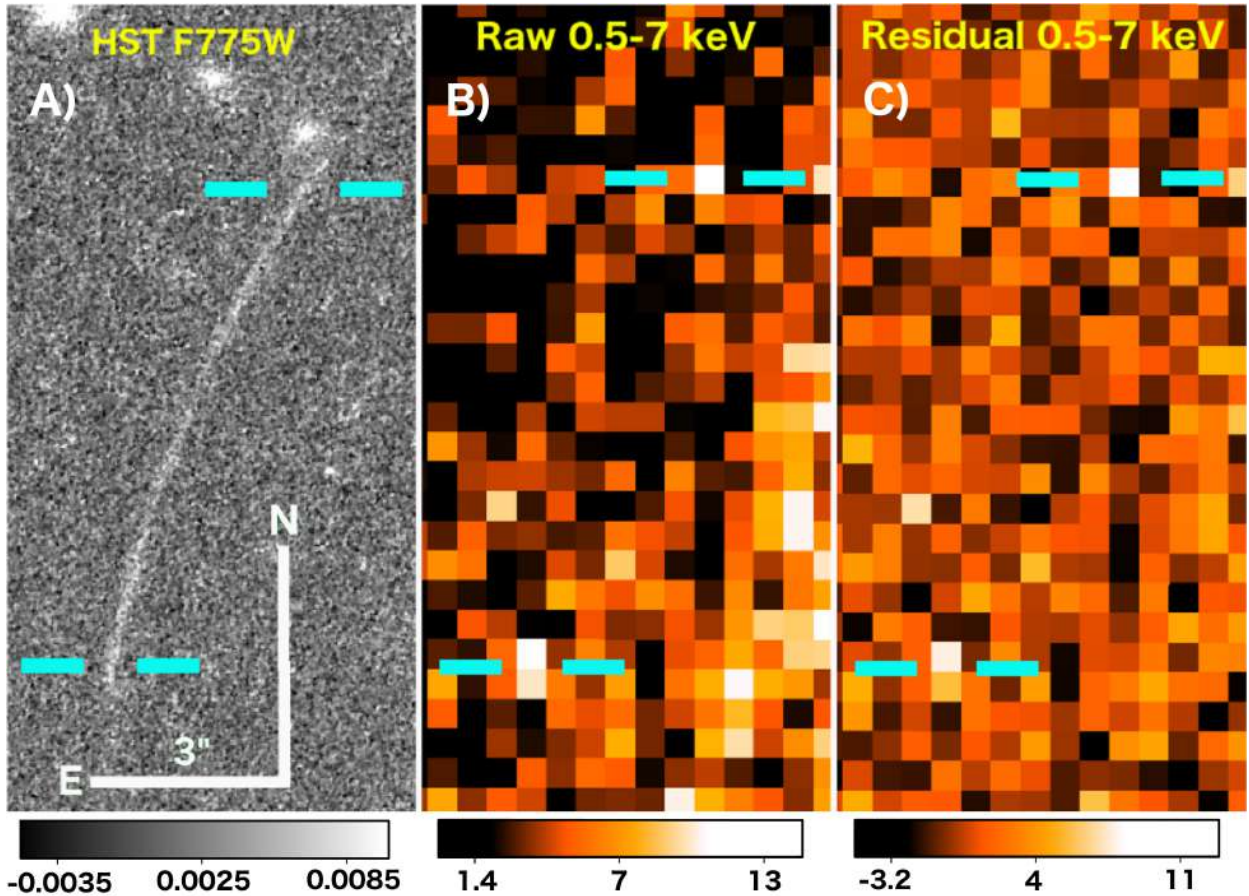

**Supplementary Figure 2 | Giant Arc X-ray Emission.** Three panels show the region of sky containing the X-ray emitting giant arc in the rest-frame UV continuum (panel A; *Hubble* F775W), the raw X-ray (panel B; 0.5-7 keV channel), and the foreground cluster-subtracted X-ray (panel C; 0.5-7 keV channel). There is X-ray emission spatially coincident with each of the ends of the giant arc, both before and after subtracting off the diffuse foreground galaxy cluster emission (locations indicated by cyan hash marks). All panels have color bars below them that indicate the mappings between pixel colors and flux values; flux values are in units of 5e-16 erg cm$^{-2}$ s$^{-1}$ Å$^{-1}$ panel A, and in 0.5-7 keV *Chandra* counts in panels B and C.

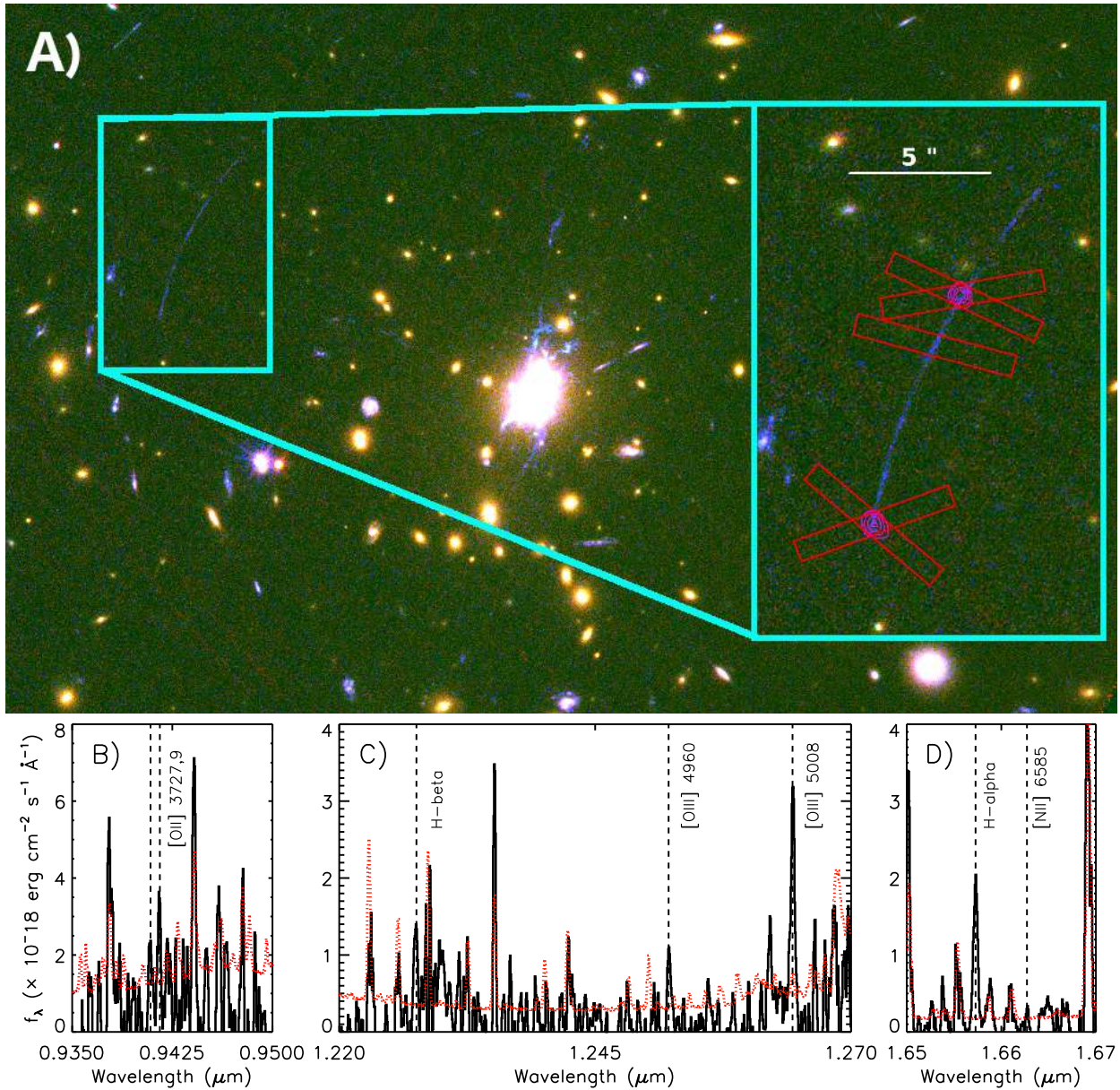

**Supplementary Figure 3 | FIRE slit positions and Emission Line Spectra.** In panel A we show the false color *Hubble* image with and inset zoom region on the X-ray emitting giant arc. The optical colors here represent *Hubble* imaging data in the F850LP (red), F775W (green), and F475W (blue) filters. In the inset we show the FIRE slit placement on the giant arc, with the positions of the multiply–imaged X-ray emission also indicated by the magenta contours; the contours are the $1/2/3/4\sigma$ levels for the individual detections of the X-ray images in a super-sampled version of the *Chandra* data. We label the two lensed images as 2.1 and 2.2, consistent with the naming scheme used below in Extended Data Figure 4. Spectra from all slit positions returned the same emission line pattern. The bottom panels show the rest–frame optical emission line spectrum of the giant arc that results from coadding the spectra obtained[7] through all of the slit positions indicated in panel A; panel B shows the [O II] 3727,3729 doublet, panel C shows H–$\beta$ and the [O III] 4960,5008 doublet, and panel D shows H–$\alpha$ and [N II] 6585.

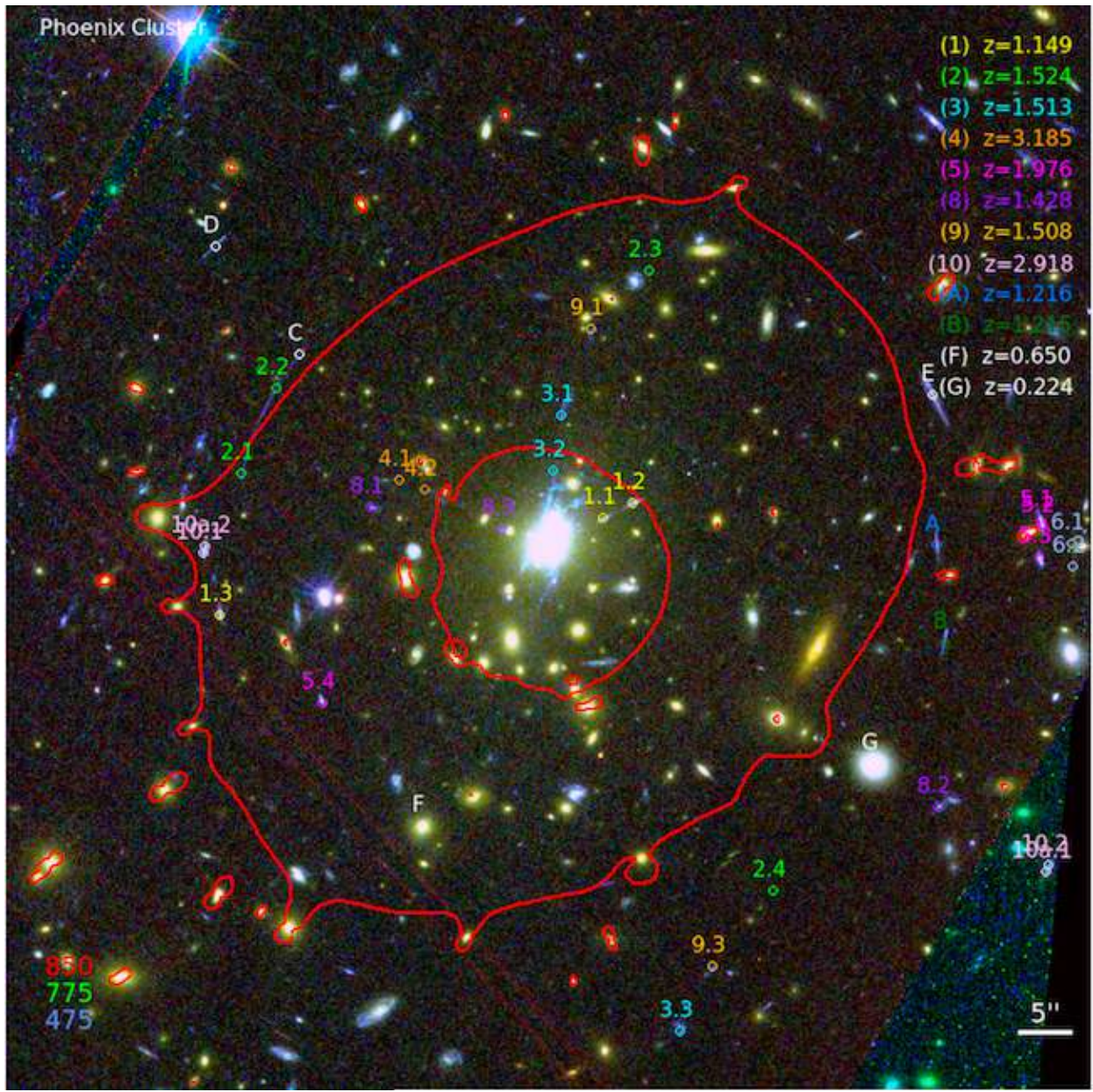

**Supplementary Figure 4 | Visualization of the Strong Lens Model.** The false color Hubble image of SPT-CLJ2344-4243 is shown here with lens model constraints indicated by numbered image families (e.g., 2.1, 2.2, 2.3 and 2.4 indicate the four lensed images of the X-ray emitting arc). Spectroscopically confirmed redshifts of lensed background sources used in the modeling are shown in the top right corner, and color-coded to match the image family markers in the image and in Supplementary Table 2. The critical curve at the redshift of the X-ray arc (z=1.5244) is also shown in red. The optical colors here are given by *Hubble* imaging data in the F850LP (red), F775W (green), and F475W (blue) filters.

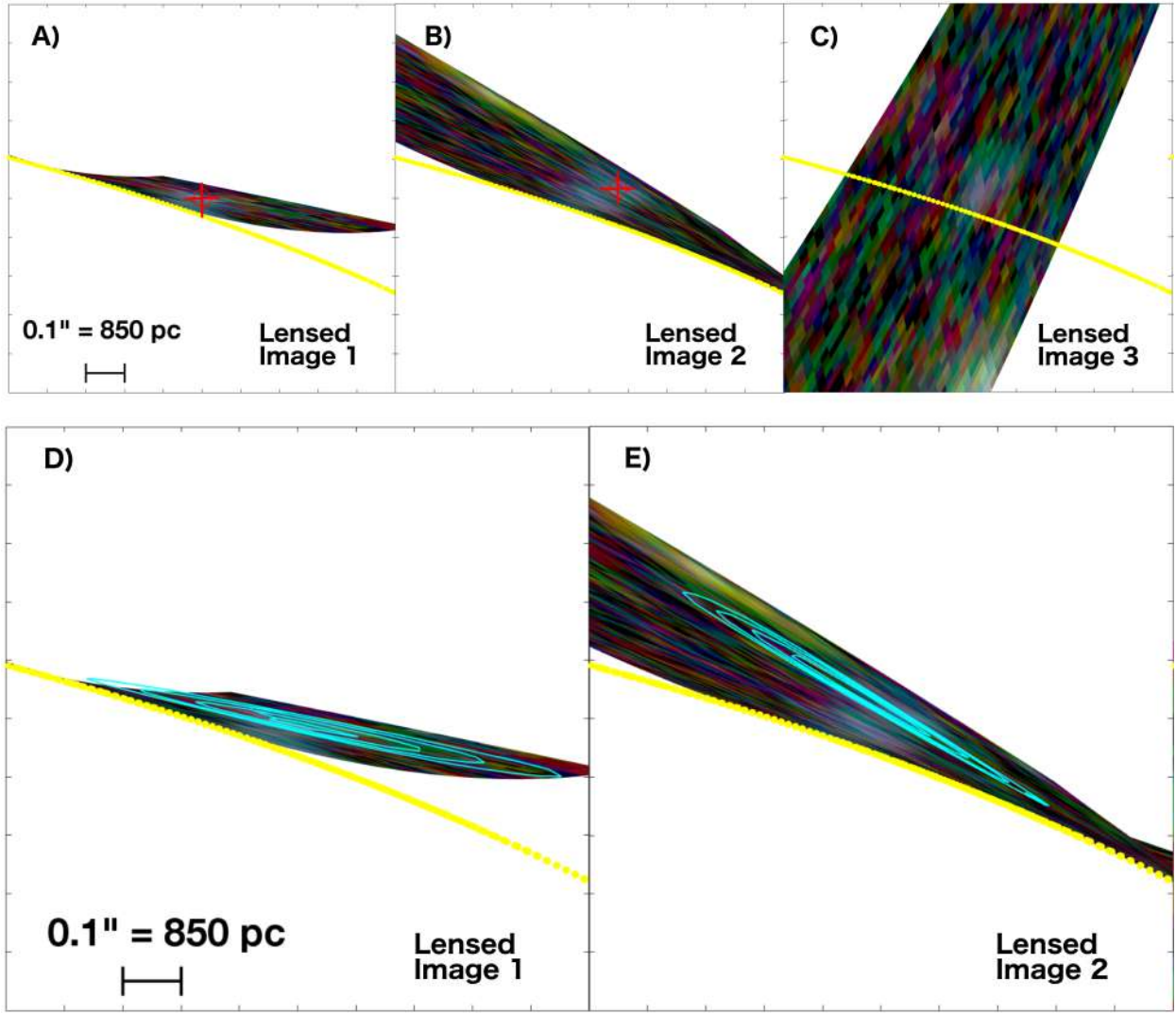

**Supplementary Figure 5 | Source Plane Reconstruction of the Giant Arc.** Here we show source–plane reconstructions of the lensed images 1 and 2 from the highly magnified giant arc with the centroid of the X-ray emission in the source–plane is indicated by the red cross in panels A and B, as well as of a less magnified—and therefore less distorted—third image (panel C). In panels D and E we show source–plane reconstructions of images 1 & 2 with the contour levels (cyan) from the top panel inset of Extended Data Figure 3 ray-traced into the source plane. The contours are the $1/2/3/4\sigma$ levels for the individual detections of the X-ray images in a super-sampled version of the *Chandra* data. These contours effectively represent the shape and extent of the *Chandra* PSF in the source plane and they show that the X-ray emitting UV–bright region in the lensed galaxy is at best marginally resolved in X-rays, thus explaining the point source nature of the detected X-ray emission. The strong lensing caustic (the critical curve mapped into the source plane) is shown as the solid yellow line in each source plane reconstructed image.

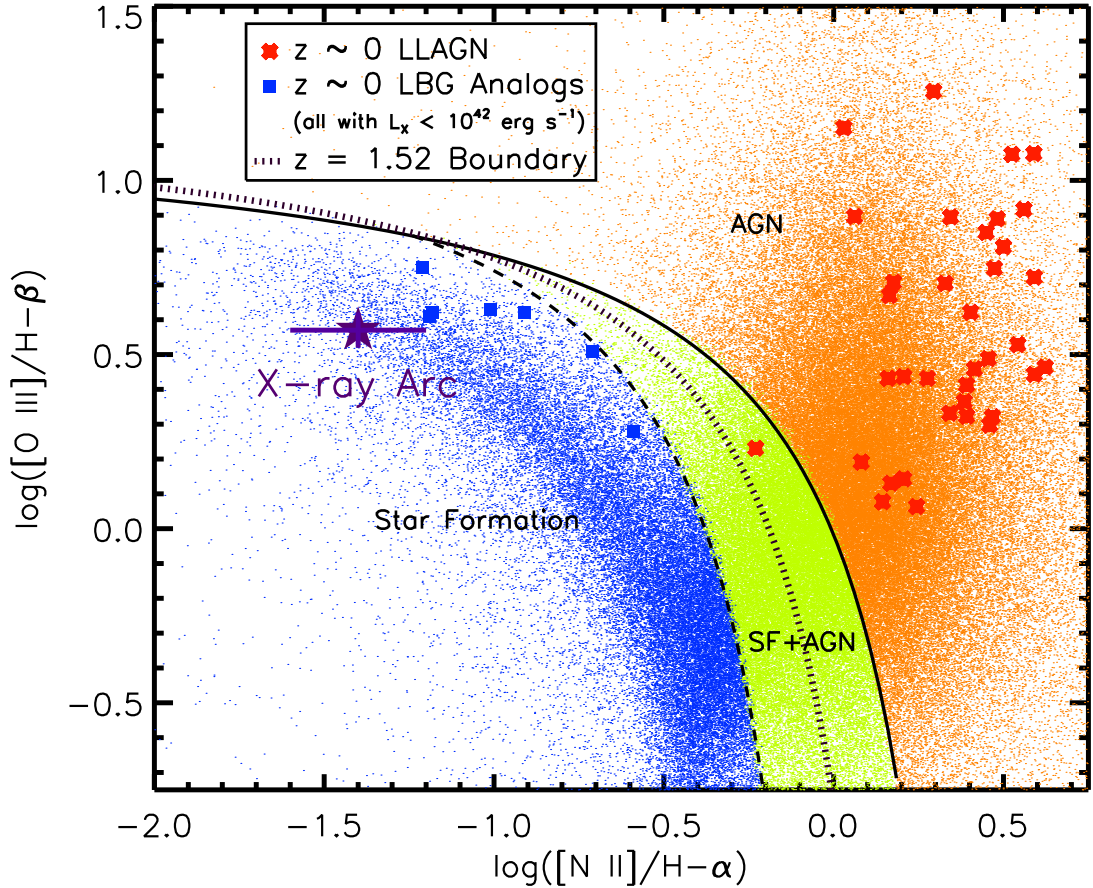

**Supplementary Figure 6 | Optical emission line ratio diagnostic diagram.** Gas in galaxies is generally ionized by either hot stars or active galactic nuclei; the ionizing source can generally be distinguished using the [O III] to H$\beta$ and [N II] to H$\alpha$ line strength ratios. Black solid and dashed lines indicate the boundaries between gas ionized by star formation, AGN, or by both SF and AGN in the local universe, while the dotted line shows the predicted evolution in the boundary between between star formation and AGN–dominated ionizing radiation at the redshift of our X-ray arc[2]. Blue, green and red clouds of points are local galaxies from the Sloan Digital Sky Survey (SDSS)[3]. Red are galaxies known to contain active galactic nuclei (AGN), blue are star-forming galaxies, and green are composite star-forming + AGN galaxies. We also use larger symbols to overplot two other comparison samples that have X-ray luminosities similar to the giant arc in SPT-CLJ2344-4243. The first, indicated by red x marks, are local low X-ray luminosity AGN (LLAGN). The second, indicated by blue filled squares, are local Lyman break galaxy (LBG) analogs, which are low-mass, UV-bright star-forming galaxies. All sources from the literature are plotted without error bars to make the plot easier to read; the literature uncertainties in both line ratios are typically <10% (i.e., <0.1 dex). The giant arc in SPT-CLJ2344-4243 is also plotted (purple filled star).

# References

1. Bayliss, M. B. *et al.* SPT-GMOS: A Gemini/GMOS-South Spectroscopic Survey of Galaxy Clusters in the SPT-SZ Survey. *Astrophys. J. Suppl.* **227**, 3 (2016).

2. Kewley, L. J. *et al.* The Cosmic BPT Diagram: Confronting Theory with Observations. *Astrophys. J. Lett.* **774**, L10 (2013).

3. Thomas, D. *et al.* Stellar velocity dispersions and emission line properties of SDSS-III/BOSS galaxies. *Mon. Not. R. Atron. Soc.* **431**, 1383–1397 (2013).